\documentclass[aps,pra,reprint,superscriptaddress,graphicx]{revtex4-1}
\usepackage{graphicx,amsfonts,xcolor}
\draft 

\begin{document}


\title{Probing structural chirality of crystals using high harmonic generation in solids} 



\author{Zi-Yu Chen}
\email[]{ziyuch@scu.edu.cn}
\affiliation{Key Laboratory of High Energy Density Physics and Technology (MoE), College of Physics, Sichuan University, Chengdu 610064, China}
\author{Rui Qin}
\email[]{qinrui.phy@outlook.com}
\affiliation{National Key Laboratory of Shock Wave and Detonation Physics, Institute of Fluid Physics, China Academy of Engineering Physics, Mianyang 621999, China}
%


\date{\today}

\begin{abstract}
Structural chirality plays an important role in solid state physics and leads to a variety of novel physics. The feasibility of probing structural chirality of crystals using high harmonic generation in solids is explored in this work. Through first-principles calculations based on the time-dependent density functional theory framework, we demonstrate that evident circular dichroism (CD) effects can be induced in the high harmonic spectra from a chiral crystal --- bulk tellurium. The CD signal reverses for crystals with opposite structural chirality. Besides, the high harmonic spectroscopy also provides an all-optical method for probing lattice symmetry properties and determining orientation of the tellurium crystal.
\end{abstract}


\maketitle 

\section{Introduction}
Chirality is a property of asymmetry in matter which can exist at every length scale and play an important role in various branches of science\cite{Riva2017}. An object is chiral if it cannot be superposed onto its mirror image, with human hands as one of the examples. The characterization of chirality has been the subject of intensive studies. 

Several chiral-sensitive spectroscopy techniques utilizing interaction with circularly polarized light have been developed for discrimination of chiral molecules. For instance, circular dichroism (CD) of photoabsorption spectroscopy in the condensed phase\cite{Fasman2013} and photoelectron circular dichroism (PECD) in the gas phase\cite{Boewering2001,Lux2012} are widely used for detecting chiral chemical and biological molecules. Besides, nonlinear optical processes in the perturbative regime have been explored to probe chirality in solution and on surfaces\cite{Fischer2005}. Recently, nonperturbative high harmonic generation (HHG) from chiral gas molecules has also been demonstrated to be chirality sensitive\cite{Cireasa2015,Smirnova2015,Baykusheva2018,Neufeld2018,Harada2018,Neufeld2019}, thus providing an attractive approach to detect chiral structures and resolve ultrafast chiral dynamics in the gas phase.

Chirality also plays a critical role in solid state physics. Out of the 230 space groups for non-magnetic materials, 65 space groups represent structurally chiral crystals\cite{Flack2003}. Structurally chirality can lead to a variety of novel physics such as multiferroicity\cite{Spaldin2005}, magnetochiral dichroism\cite{Fasman2013}, skyrmions\cite{Muehlbauer2009}, current-induced magnetizations\cite{Yoda2015},
circular photogalvanic effect\cite{Tsirkin2018}, and universal topological electronic properties\cite{Chang2018,Sanchez2019}. Thus it is of great interest to detect and characterize the structural chirality embedded in crystals.

The recent advent of HHG in solids\cite{Ghimire2011,Ghimire2019} has opened up exciting opportunities for not only the development of novel solid-state extreme-ultraviolet and attosecond photonics with bulk\cite{TD2017b,Langer2017,Hammond2017,Sivis2017,Garg2018} or at nanoscale\cite{Yoshikawa2017,Taucer2017,Baudisch2018,chen_circularly_2019,Qin2018,Chen_BP_2019,guan_cooperative_2019,guan_toward_2020,yoshikawa_interband_2019,TD2018,le_breton_high-harmonic_2018}, but also the investigation of strong-field and ultrafast dynamics in the condensed phase using the high harmonic spectroscopy\cite{Vampa2014,Luu2015,TD2017a,klemke_polarization-state-resolved_2019}. Substantial progress has been reported, such as the demonstration of probing atomic-scale crystal structure\cite{You2017}, reconstructing electronic band structure\cite{Vampa2015b,lanin_mapping_2017}, measuring Berry curvature\cite{Luu2018}, characterizing complex ultrafast many-body dynamics in strongly-correlated systems\cite{silva_high-harmonic_2018}, and revealing strong-field physics in topological systems\cite{bauer_high-harmonic_2018,silva_topological_2019}. Like its counterpart in chiral gas harmonics, chiral HHG in solids may also be useful in revealing chiral structures and dynamics in the condensed phase.

In this work, we extends the chiral high harmonic spectroscopy to solids. The feasibility of probing structural chirality of crystals using HHG in solids is explored. Through first-principles calculations based on the time-dependent density functional theory (TDDFT) framework, we demonstrate that considerable CD signals in the HHG spectra can be induced when a crystal with chiral lattice interacts with circularly polarized light. In addition, the high harmonic spectroscopy also provides an optical method to probe lattice symmetry properties and determine orientation of the crystal.

\section{Crystal structure}

\begin{figure}[htbp]
\centering
\includegraphics[width=0.48\textwidth
]{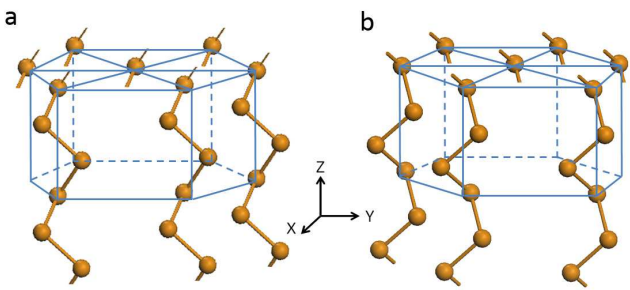}
\caption{\label{structure} Crystal structure of bulk tellurium with space groups (\textbf{a}) $P3_221$ (\#154 left-handed) and (\textbf{b}) $P3_121$ (\#152 right-handed). Helical tellurium atomic chains arranged in a hexagonal array spiraling around axes parallel to the $z$-axis. The screw structure of the atoms along the spiral axis for each space group is in the opposite direction.}
\end{figure}

We choose the simplest material with a chiral structure, i.e., elemental tellurium, to study the structural chirality in crystals. Tellurium has a trigonal crystal structure with three atoms in the unit cell. The unique structure consists of helical chains arranged in a hexagonal array spiraling around axes parallel to the $z$-axis ([0001] direction) at the corners and center of the hexagonal elementary cell, as shown in Fig.~\ref{structure}. Depending on the screw direction of the helical chains, there are two enantiomorphic crystal structures with space groups $P3_221$ (\#154 left-handed) and $P3_121$ (\#152 right-handed)\cite{Asendorf1957}. Each tellurium atom is covalently bonded with its two nearest neighbors on the same chain, while forms weak van der Waals-like interchain bonds with its four next nearest neighbors. Thus there exists inherent structural anisotropy in the tellurium crystal\cite{Peng2015}.

\begin{figure*}[htbp]
\centering
\includegraphics[width=0.98\textwidth
]{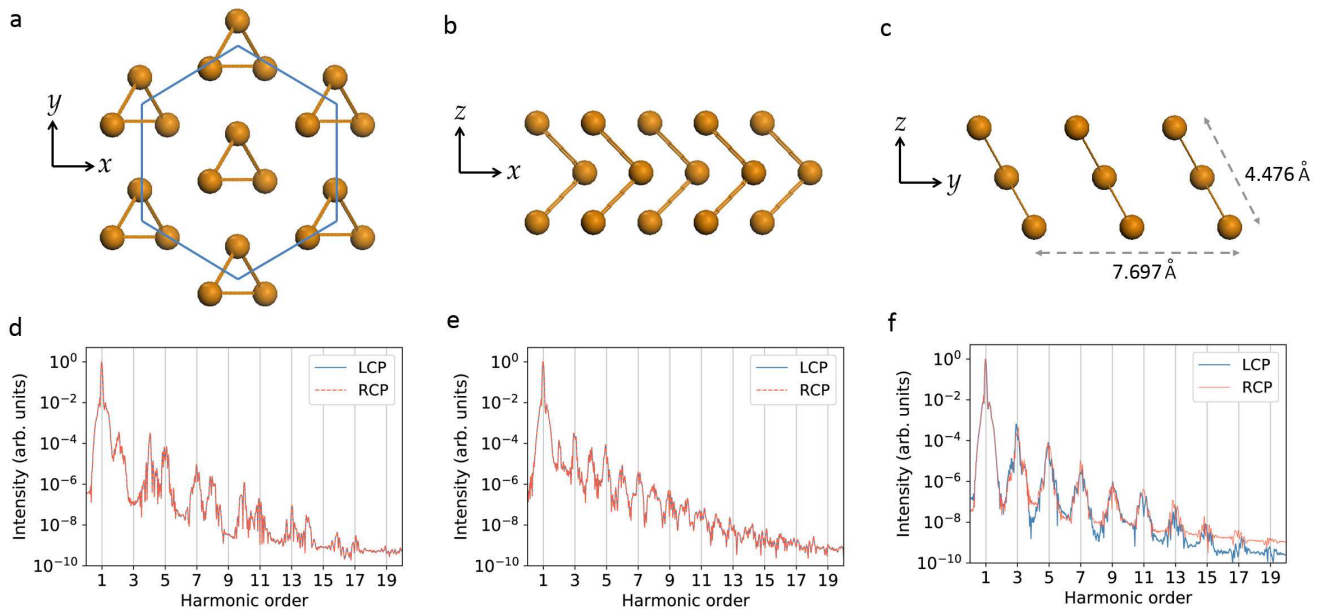}
\caption{\label{symmetry} Orientation-dependent high harmonic generation in the tellurium crystal with the $P3_121$ space group. (\textbf{a}-\textbf{c}) Crystal structures in the $x$-$y$ plane (panel \textbf{a}; top view), $x$-$z$ plane (panel \textbf{b}; side view), and $y$-$z$ plane (panel \textbf{c}; side view). (\textbf{d}-\textbf{f}) High harmonic spectra for different crystal orientations corresponding to (\textbf{a}-\textbf{c}), respectively. For each crystal orientation, circularly polarized lasers of opposite handedness are incident onto the crystal under normal incidence. LCP and RCP denote the recorded high harmonics driven by left- and right-handed circularly polarized laser pulses, respectively. The presence of harmonic orders reflects the crystalline symmetries in each plane. The laser wavelength is 800 nm and intensity is $8\times 10^{11}$ W/cm$^2$.}
\end{figure*}

\section{Computational method}

The geometric structure of tellurium crystal is relaxed by using the CASTEP package\cite{castep2005} within the density functional theory framework. The ultrasoft pseudopotentials are used in the calculations, and the plane-wave cutoff energy is set to be 210 eV. The generalized gradient approximation (GGA) of the PBE form\cite{Perdew1996} is employed for the exchange-correlation functional, and van der Waals interaction correction is considered  by using the Tkatchenko and Scheffler method\cite{Tkatchenko2009}. 

Time evolution of the wave functions and time-dependent electronic current are studied by using the OCTOPUS package\cite{Andrade2015}, where the Kohn-Sham equations are propagated in real time and real space within the TDDFT framework in the adiabatic local density approximation. We have also tested using a different exchange-correlation functional, i.e., adiabatic GGA (PBE), with pseudopotentials given by Ref.\cite{schlipf_optimization_2015}, which gives similar results. In all the calculations, the real-space spacing is 0.4 Bohr. The fully relativistic Hartwigsen, Goedecker, and Hutter pseudopotentials are used. A $24\times 24 \times 16$ Monkhorst-Pack k-point mesh is used for Brillouin zone sampling. 

The laser field is described in the velocity gauge. The vector potential of the circularly polarized laser has the following form:
\begin{equation}
\textbf{A}(t)=\frac{\sqrt{I_0 }c}{\omega}f(t)\Big[ \frac{\cos(\omega t +\phi)}{\sqrt{1+\epsilon^2}} \hat{\textbf{e}}_i + \frac{\epsilon \sin(\omega t +\phi)}{\sqrt{1+\epsilon^2}} \hat{\textbf{e}}_j \Big],
\end{equation}
where $I_0=8\times 10^{11}$ W/cm$^2$ is the laser peak intensity in vacuum, $f(t)=\sin^2(\pi t/2\tau)$ is the sin-squared pulse envelope with $ \tau=20 $ fs, $\omega$ is the laser photon frequency, $\phi$ is the carrier-envelop phase, $\epsilon$ is the laser ellipticity, $c$ is the light speed in vacuum, $\hat{\textbf{e}}_i$ and $\hat{\textbf{e}}_j$ with $(i,j)\in \{(x,y),(x,z),(y,z)\}$ are the unit vectors, respectively.
We use Ti:sapphire laser pulses with wavelength of $\lambda_L=$ 800 nm, corresponding to $\omega=1.55$ eV. The carrier-envelope phase is taken to be $\phi= 0$.
Left-handed circularly polarized (LCP) and right-handed circularly polarized (RCP) laser pulses are considered in this work, corresponding to $\epsilon=-1$ and $\epsilon=1$, respectively.

The HHG spectrum is calculated from the time-dependent electronic current $\textbf{j}(\textbf{r},t)$ as:
\begin{equation}
\mathrm{HHG}(\omega) = \Big| \mathcal{FT} \Big(\frac{\partial}{\partial t} \int \textbf{j}(\textbf{r},t) \ \mathrm{d}^3 \textbf{r}  \Big) \Big|^2,
\end{equation}
where $\mathcal{FT}$ denotes the Fourier transform. 

The total number of excited electrons is calculated by projecting the time-dependent Kohn-Sham states onto the ground-state Kohn-Sham states. As the \textit{n}-th state evolves in time, it has some possibility to transit to other states and thus contains other ground-state components. The total number of excited electrons $N_{ex}(t)$ is calculated as:
\begin{equation}
N_{ex}(t)=N_e-\int_{\mathrm{BZ}} \sum_{n,m}^{\mathrm{occ}} f_{n,k} |\langle\phi_{n,k}(t)|\phi_{m,k}(0)\rangle|^2 \mathrm{d} \textbf{k}
\end{equation}
where $N_e$ is the total number of electrons in the system, $\phi_{n,k}(t)$ is the time-dependent Kohn-Sham state at \textit{n}-th band at k-point \textit{k}, $\phi_{m,k}(0)$ is the ground-state (t = 0) Kohn-Sham state at \textit{m}-th band at k-point \textit{k}, $f_{n,k}$ is the occupation of Kohn-Sham state at \textit{n}-th band at k-point \textit{k}, and BZ denotes integration over the whole Brillouin zone.

\section{Crystalline symmetries}

The top view (in the $x$-$y$ plane) of the crystal structure of bulk tellurium with the space group $P3_121$ (right handed) is shown in Fig.~\ref{symmetry}(a), while the side views in the $x$-$z$ and $y$-$z$ planes are shown in Fig. \ref{symmetry}(b)-\ref{symmetry}(c), respectively. In a serials of simulations, we use LCP or RCP laser pulse propagating along the $z$-, $y$-, or $x$-directions, i.e., with laser field components in the $x$-$y$, $x$-$z$, or $y$-$z$ planes. The \textit{in-plane} HHG spectra are shown in Figs. \ref{symmetry}(d)-\ref{symmetry}(f), corresponding to Figs.~\ref{symmetry}(a)-\ref{symmetry}(c), respectively. Different features of appearing harmonic orders are observed, which reflect different crystalline symmetries in each plane. In the $x$-$y$ plane, the crystal structure exhibits a three-fold rotational symmetry (Fig.~\ref{symmetry}(a)). Since the circularly polarized laser field can be assumed isotropic in the $x$-$y$ polarization plane, the laser-crystal interaction system thus possesses a three-fold rotational symmetry in this plane. For such a system, the harmonic order $n_\mathrm{H}$ obeys a simple selection rule of $n_\mathrm{H}=3m\pm 1$ ($m \in \mathbb{N}$), i.e., every third harmonic is suppressed\cite{Saito2017,Chen2018,chen_circularly_2019}. The corresponding harmonic spectra shown in Fig.~\ref{symmetry}(d) agree well with this selection rule. In the $x$-$z$ plane, the crystal structure lacks a inversion symmetry (Fig.~\ref{symmetry}(b)). As a result, both odd and even harmonics are present in the corresponding harmonic spectra (Fig.~\ref{symmetry}(e)). In contrast, the projected crystal structure in the $y$-$z$ plane is centrosymmetric (Fig.~\ref{symmetry}(c)). Consequently, only odd harmonics can be observed in the spectra (Fig.~\ref{symmetry}(f)). Therefore, the feature of allowed harmonic orders of the HHG spectra can be employed to determine the lattice symmetry and crystal orientation of bulk-tellurium.

\section{Structural chirality}
\begin{figure}[htbp]
\centering
\includegraphics[width=0.5\textwidth
]{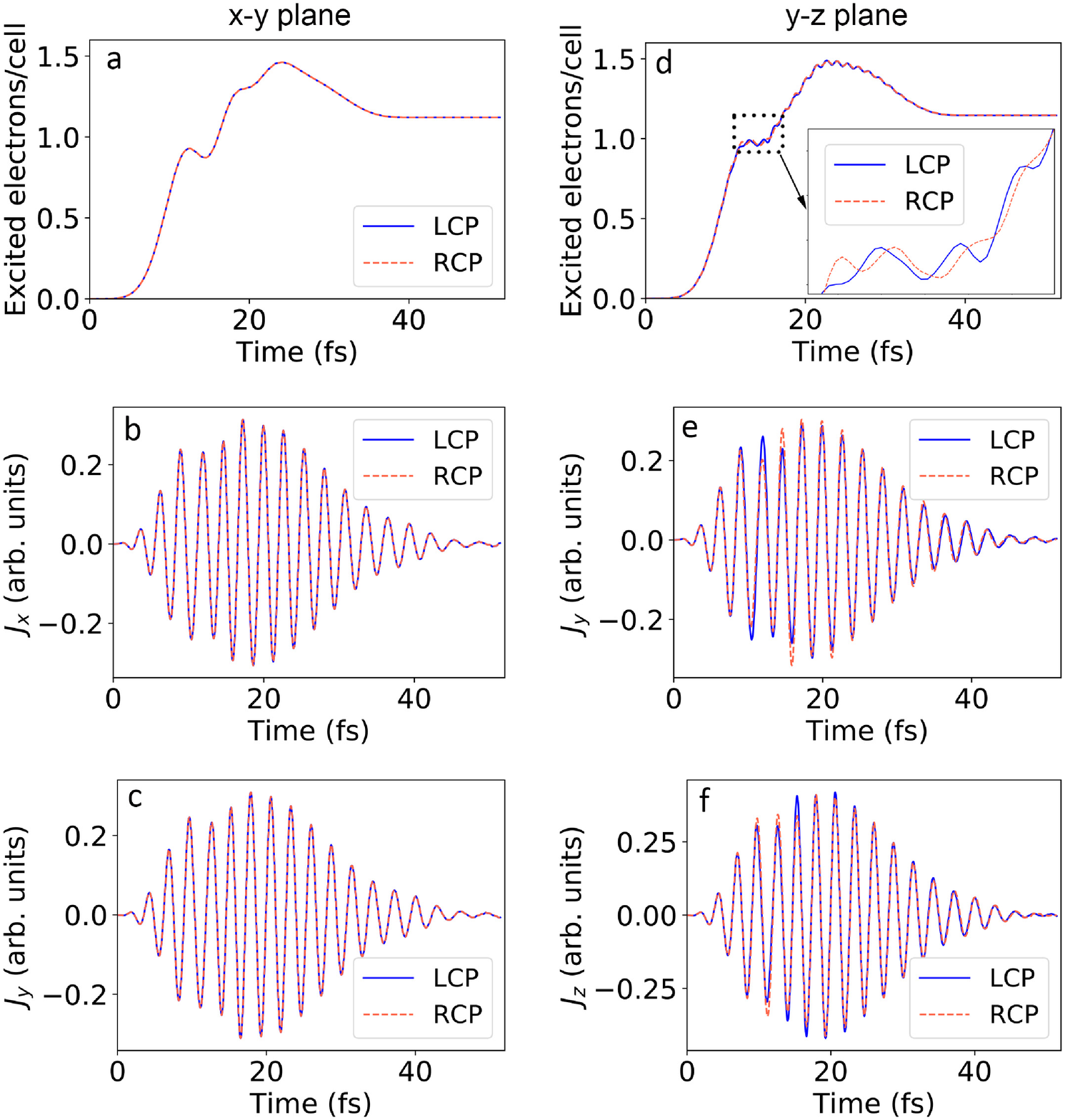}
\caption{\label{occ} Circular dichroism effect on the dynamics of electron excitation and electronic current. (\textbf{a}, \textbf{d}) Time evolution of the number of electrons excited to the conduction bands and (\textbf{b}-\textbf{f}) electronic current during the laser pulse. Panels (\textbf{a}-\textbf{c}) are for laser polarization and crystal structure in the $x$-$y$ plane, while panels (\textbf{d}-\textbf{f}) are for the $y$-$z$ plane. The laser wavelength is 800 nm and intensity is $8\times 10^{11}$ W/cm$^2$.}
\end{figure}

Apart from the harmonic orders, another important characteristic of the HHG spectra shown in Figs.~\ref{symmetry}(d)-\ref{symmetry}(f) is the different harmonic-intensity response to the LCP and RCP lasers for different crystal orientations and related lattice structures. While the LCP and RCP laser-driving HHG intensity show no difference in Figs.~\ref{symmetry}(d)-\ref{symmetry}(e), appreciable CD effect over a broad range of harmonic orders can be observed in the HHG spectra shown in Fig. \ref{symmetry}(f). This CD effect can also be seen clearly from the time evolution of the excited electron numbers and total electronic currents. For the case of $x$-$y$ plane, either the number of excited electrons (Fig.~\ref{occ}(a)) or the total currents (Figs.~\ref{occ}(b-c)) shows no difference between LCP and RCP laser pumps. In comparison, for the case of $y$-$z$ plane, the dynamics of carrier excitation to the conduction bands during the laser pulse show different responses to LCP and RCP driving lasers in both excitation amplitude and phases (Fig. \ref{occ}(d)). This means interacting with such a crystal structure (Fig.~\ref{symmetry}(c)) lasers of opposite handedness can have different dynamic absorption and ionization probabilities. While the CD effect on the excitation response is small, a pronounced difference can be seen in the total current (Figs.~\ref{occ}(e-f)).
 
The CD effect can be explained by the crystal's structural chirality-induced chiroptical response. Mathematically, a figure is said to have chirality if it cannot be mapped to its mirror image by (proper) rotations and translations alone. As a result, a chiral figure cannot posses an axis of symmetry in 2D. Based on this criterion, it is easy to see that the crystal structures in the $x$-$y$ and $x$-$z$ planes (Figs.~\ref{symmetry}(a)-\ref{symmetry}(b)) are achiral, while the crystal structure in the $y$-$z$ plane (Fig.~\ref{symmetry}(c)) indeed posses chirality. As circularly polarized lasers are chiral objects exhibiting optical chirality, in their interaction with another chiral object, i.e., the chiral crystal, LCP and RCP laser-induced HHG response become distinguishable. 

In the velocity gauge, the general Hamiltonian can be written as $\hat{H} = \sum_j\left[(\textit{\textbf{p}}_j+\textit{\textbf{A}}_j)^2/2 + V_j\right]+W$, where $\textit{\textbf{p}}_j=-i\nabla$ is the canonical momentum operator for the j-th electron, $\textit{\textbf{A}}$ is the (chiral) electromagnetic field vector potential, $V$ is the (chiral) Coulomb scalar potential, and $W$ is the interaction between electrons. In this gauge, both electric dipole and non-dipole interaction effects are accounted for the HHG. As atoms are oriented in crystals, here electric dipole effect is likely to dominate over non-dipole effects in the chiral contribution\cite{Mairesse2008,Wang2017}.

\begin{figure}[htbp]
\centering
\includegraphics[width=0.35\textwidth
]{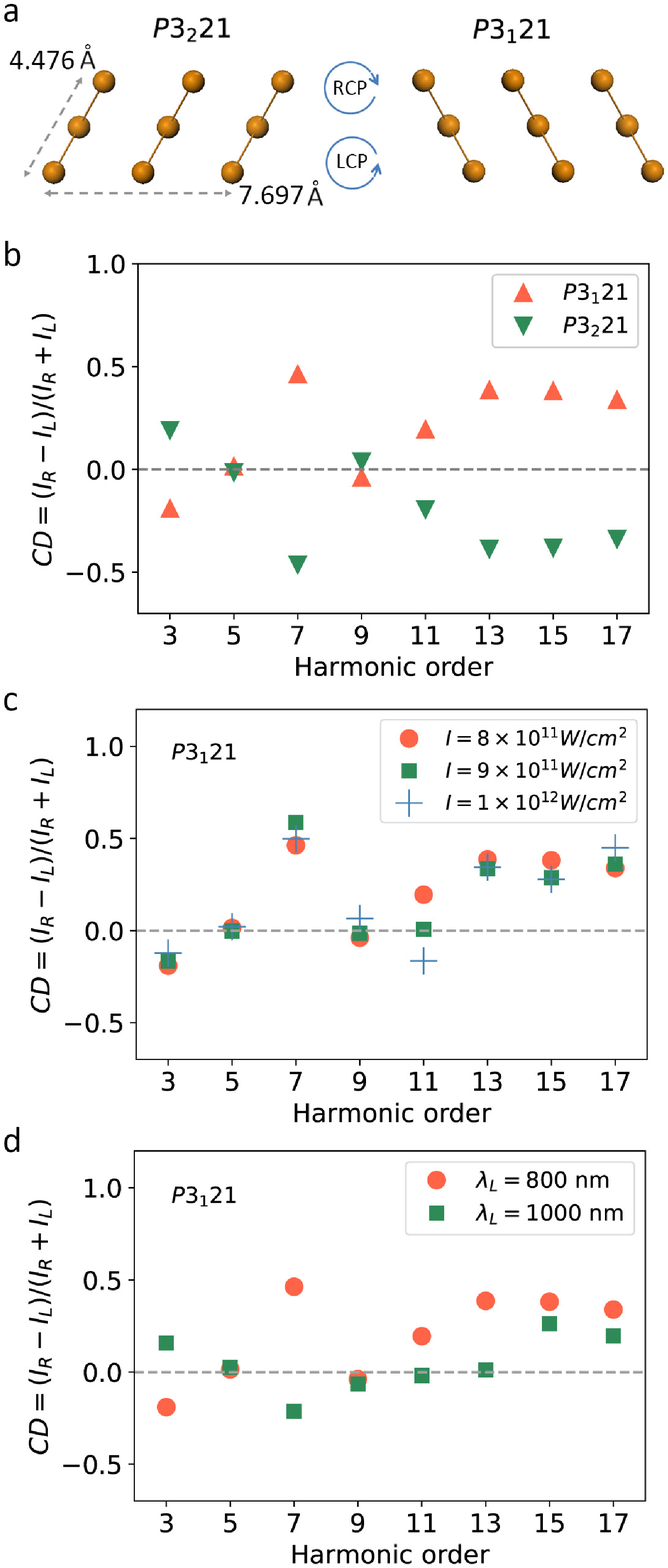}
\caption{\label{chiral} Chiral response of high harmonic emission. (\textbf{a}) Two enantiomorphic crystal structures of tellurium with space groups $P3_121$ (right handed) and $P3_221$ (left handed) in the $y$-$z$ plane. Left- and right-handed circularly polarized (LCP \& RCP) lasers are also in the same plane. (\textbf{b}) Quantitatively evaluated circular dichroism (CD) signal from the enantiomers for each harmonic order. The laser wavelength is 800 nm and intensity is $8\times 10^{11}$ W/cm$^2$. CD signal as a function of harmonic photon energy for different (\textbf{c}) laser intensity (with wavelength of 800 nm) and (\textbf{d}) laser wavelength (with intensity of $8\times 10^{11}$ W/cm$^2$).}
\end{figure}

As the applied laser wavelength is much larger than the relevant atomic dimensions, long-wavelength approximation is adopted in the TDDFT calculations. As such, any spatial dependence of the field is neglected in the simulations, i.e., $\textit{\textbf{A}}(\textit{\textbf{r}},t)\rightarrow \textit{\textbf{A}}(t)$. Then the field is spatially uniform at any time instant rather than really propagating along the wave vector direction that would require $\textit{\textbf{A}}(\omega t - \textit{\textbf{k}}\cdot \textit{\textbf{r}})$ satisfying the wave equation and depending on both space and time coordinates. This means that the field vector of the circularly polarized light in the simulations, instead of screwing in the 3D space, rotates only in the polarization planes, which are 2D planes perpendicular to the propagation axis. Consequently, the aforementioned chiroptical response is limited to this 2D effect. Nevertheless, the above results unambiguously demonstrate that 2D structural chirality can be probed using circular dichroism of HHG in solids. Therefore, the chiral HHG spectroscopy is directly applicable to study atomically thin 2D chiral crystals such as tellurene\cite{Wu2018}. Moreover, the results also imply crystal's 3D structural chirality can be probed by this method. Thus, CD signals of HHG in tellurium crystals due to the helical lattice structure along the $z$-axis can be expected in real experiments. This may provide an alternative approach to identify the handedness of elemental crystals such as tellurium or selenium\cite{Tanaka2010}.

\begin{figure}[htbp]
\centering
\includegraphics[width=0.5\textwidth
]{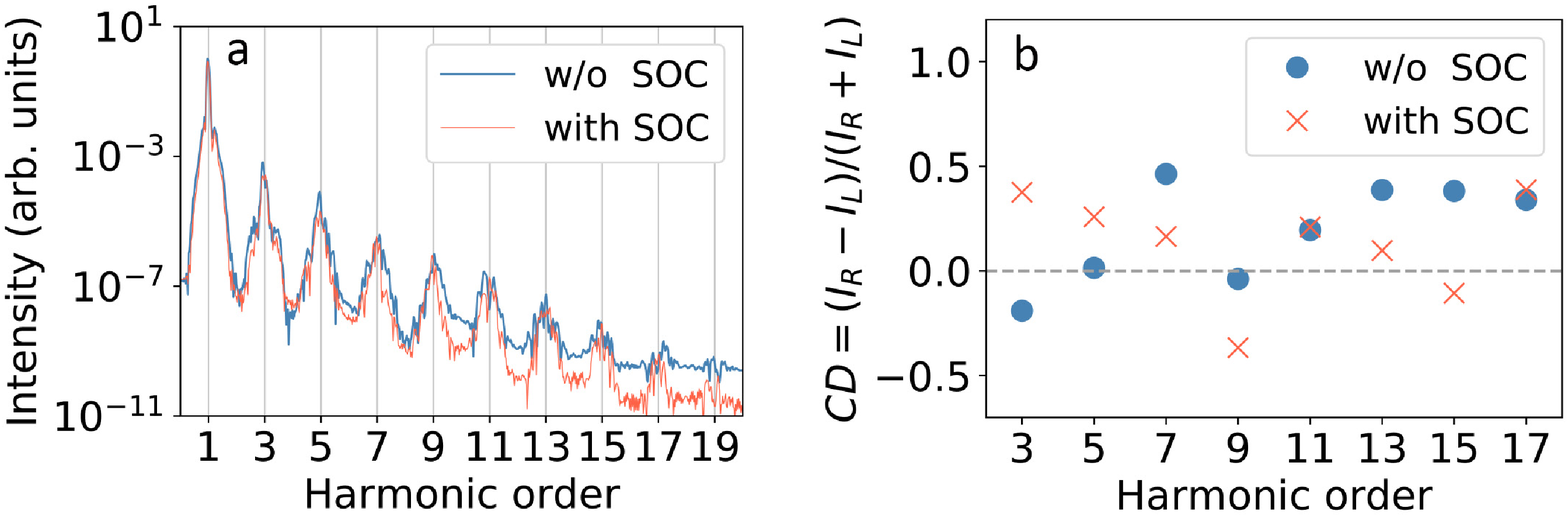}
\caption{\label{soc} Spin-orbit coupling (SOC) effect on HHG. (a) High harmonic spectra and (b) CD signals obtained with (red) and without (blue) considering SOC effect. The other simulation parameters are the same with those in Fig.~\ref{symmetry}(f).}
\end{figure}

To further demonstrate the chiroptical response can be attributed to the structural chirality, we compute the CD signals from crystal structures of opposite handedness. Figure~\ref{chiral}(a) shows the atomic configurations of tellurium crystal with the $P3_121$ (right handed) and $P3_221$ (left handed) space groups in the $y$-$z$ plane. Obviously, they are non-superimposable mirror images to each other. The quantitative CD signal is defined as
\begin{equation}\label{key}
CD = \frac{I_R-I_L}{I_R+I_L},
\end{equation}
where $I_R$ and $I_L$ denote the peak harmonic intensity for each harmonic order driven by the RCP and LCP lasers, respectively. The obtained CD signal for different harmonic orders is shown in Fig.~\ref{chiral}(b). The CD asymmetry reaches up to 50\%, much larger than that in molecules\cite{Cireasa2015,Baykusheva2018}. When the crystal structure is right-handed, more harmonic orders show stronger intensity driven by RCP lasers than by LCP lasers. Switching the handedness of crystal reverses this effect. The CD signal shows a perfect mirroring between the two crystals of opposite handedness, proving that the chiroptical response is indeed due to the structural chirality of the tellurium crystal. 

We also plot a set of CD signal data recorded at different driving laser intensities in Fig. \ref{chiral}(c). We see the change of laser intensity only has a small impact on the CD signals, in contrast to previous studies on chiral HHG from gas molecules where CD signals show substantial changes with a minor change of laser intensity\cite{Baykusheva2018}. This result suggests in this parameter regime chiral electronic dynamics taking place during the HHG process having minor contribution to the observed chiral signals. Instead, effects originating from the static structure of the crystals dominate the chiroptical response. In comparison, the dependence of CD signal on laser wavelength is relatively more sensitive, as shown in Fig. \ref{chiral}(d), implying an increased contribution from electronic dynamics. Thus both static structure and dynamic effects can contribute to the CD signal.

Finally, we address the effect of spin-orbit coupling (SOC). Previous studies suggest that SOC effect could play an important role in the electronic properties of tellurium and trigger exotic electromagnetic effects\cite{sakano_radial_2020,rodriguez_two_2020}. We have carried out additional simulations with SOC effect included and other parameters the same as those in Fig.~\ref{symmetry}(f). Figure~\ref{soc}(a) shows the peak harmonic intensity is lowered as compared to that without SOC, while the presence of harmonic orders remains the same since it is determined by the lattice symmetry. Figure~\ref{soc}(b) shows significant CD signal is still induced when SOC effect is considered, albeit may with a different value for each harmonic order. As in the present work we focus on lattice structural chirality, the SOC effect would not change our main conclusion qualitatively that structural chirality of crystals can lead to CD effect in HHG from the material. Studying influence of electronic and spin structures in tellurium on HHG would be an interesting future work. 


\section{Conclusions}
In conclusion, we demonstrate through \textit{ab initio} TDDFT calculations that chiral HHG can be induced from interaction of circularly polarized lasers with chiral crystals. While the allowed harmonic orders reflect the lattice symmetries, CD signals of harmonic intensity shed light on the structural chirality. It is worth noting that there exists different phases for 2D tellurene (atomically thin tellurium), e.g., the $\alpha$-, $\beta$-, $\gamma$-, and $\delta$-phase\cite{Xiang2019}. As the different phases of tellurene exhibit different symmetries and chirality, the circularly polarized HHG spectroscopy is also potentially useful for characterizing the phases and phase transition dynamics of this emerging 2D material.


\section*{acknowledgments}
This work was supported in part by the National Natural Science Foundation of China (Grant No. 11705185), the Presidential Fund of China Academy of Engineering Physics (Grant No. YZJJLX2017002), and the Fundamental Research Funds for the Central Universities.

\bibliography{ref_chiral}

\begin{thebibliography}{63}%
\makeatletter
\providecommand \@ifxundefined [1]{%
 \@ifx{#1\undefined}
}%
\providecommand \@ifnum [1]{%
 \ifnum #1\expandafter \@firstoftwo
 \else \expandafter \@secondoftwo
 \fi
}%
\providecommand \@ifx [1]{%
 \ifx #1\expandafter \@firstoftwo
 \else \expandafter \@secondoftwo
 \fi
}%
\providecommand \natexlab [1]{#1}%
\providecommand \enquote  [1]{``#1''}%
\providecommand \bibnamefont  [1]{#1}%
\providecommand \bibfnamefont [1]{#1}%
\providecommand \citenamefont [1]{#1}%
\providecommand \href@noop [0]{\@secondoftwo}%
\providecommand \href [0]{\begingroup \@sanitize@url \@href}%
\providecommand \@href[1]{\@@startlink{#1}\@@href}%
\providecommand \@@href[1]{\endgroup#1\@@endlink}%
\providecommand \@sanitize@url [0]{\catcode `\\12\catcode `\$12\catcode
  `\&12\catcode `\#12\catcode `\^12\catcode `\_12\catcode `\%12\relax}%
\providecommand \@@startlink[1]{}%
\providecommand \@@endlink[0]{}%
\providecommand \url  [0]{\begingroup\@sanitize@url \@url }%
\providecommand \@url [1]{\endgroup\@href {#1}{\urlprefix }}%
\providecommand \urlprefix  [0]{URL }%
\providecommand \Eprint [0]{\href }%
\providecommand \doibase [0]{https://doi.org/}%
\providecommand \selectlanguage [0]{\@gobble}%
\providecommand \bibinfo  [0]{\@secondoftwo}%
\providecommand \bibfield  [0]{\@secondoftwo}%
\providecommand \translation [1]{[#1]}%
\providecommand \BibitemOpen [0]{}%
\providecommand \bibitemStop [0]{}%
\providecommand \bibitemNoStop [0]{.\EOS\space}%
\providecommand \EOS [0]{\spacefactor3000\relax}%
\providecommand \BibitemShut  [1]{\csname bibitem#1\endcsname}%
\let\auto@bib@innerbib\@empty
\bibitem [{\citenamefont {Riva}(2017)}]{Riva2017}%
  \BibitemOpen
  \bibfield  {author} {\bibinfo {author} {\bibfnamefont {S.}~\bibnamefont
  {Riva}},\ }\bibfield  {title} {\bibinfo {title} {Chirality in metals: an
  asymmetrical journey among advanced functional materials},\ }\href@noop {}
  {\bibfield  {journal} {\bibinfo  {journal} {Mater. Sci. Technol.}\ }\textbf
  {\bibinfo {volume} {33}},\ \bibinfo {pages} {795} (\bibinfo {year}
  {2017})}\BibitemShut {NoStop}%
\bibitem [{Fas(2013)}]{Fasman2013}%
  \BibitemOpen
  \href@noop {} {\emph {\bibinfo {title} {Circular Dichroism and the
  Conformational Analysis of Biomolecules}}}\ (\bibinfo  {publisher}
  {Springer},\ \bibinfo {year} {2013})\BibitemShut {NoStop}%
\bibitem [{\citenamefont {B\"owering}\ \emph {et~al.}(2001)\citenamefont
  {B\"owering}, \citenamefont {Lischke}, \citenamefont {Schmidtke},
  \citenamefont {M\"uller}, \citenamefont {Khalil},\ and\ \citenamefont
  {Heinzmann}}]{Boewering2001}%
  \BibitemOpen
  \bibfield  {author} {\bibinfo {author} {\bibfnamefont {N.}~\bibnamefont
  {B\"owering}}, \bibinfo {author} {\bibfnamefont {T.}~\bibnamefont {Lischke}},
  \bibinfo {author} {\bibfnamefont {B.}~\bibnamefont {Schmidtke}}, \bibinfo
  {author} {\bibfnamefont {N.}~\bibnamefont {M\"uller}}, \bibinfo {author}
  {\bibfnamefont {T.}~\bibnamefont {Khalil}},\ and\ \bibinfo {author}
  {\bibfnamefont {U.}~\bibnamefont {Heinzmann}},\ }\bibfield  {title} {\bibinfo
  {title} {Asymmetry in photoelectron emission from chiral molecules induced by
  circularly polarized light},\ }\href@noop {} {\bibfield  {journal} {\bibinfo
  {journal} {Phys. Rev. Lett.}\ }\textbf {\bibinfo {volume} {86}},\ \bibinfo
  {pages} {1187} (\bibinfo {year} {2001})}\BibitemShut {NoStop}%
\bibitem [{\citenamefont {Lux}\ \emph {et~al.}(2012)\citenamefont {Lux},
  \citenamefont {Wollenhaupt}, \citenamefont {Bolze}, \citenamefont {Liang},
  \citenamefont {K\"ohler}, \citenamefont {Sarpe},\ and\ \citenamefont
  {Baumert}}]{Lux2012}%
  \BibitemOpen
  \bibfield  {author} {\bibinfo {author} {\bibfnamefont {C.}~\bibnamefont
  {Lux}}, \bibinfo {author} {\bibfnamefont {M.}~\bibnamefont {Wollenhaupt}},
  \bibinfo {author} {\bibfnamefont {T.}~\bibnamefont {Bolze}}, \bibinfo
  {author} {\bibfnamefont {Q.}~\bibnamefont {Liang}}, \bibinfo {author}
  {\bibfnamefont {J.}~\bibnamefont {K\"ohler}}, \bibinfo {author}
  {\bibfnamefont {C.}~\bibnamefont {Sarpe}},\ and\ \bibinfo {author}
  {\bibfnamefont {T.}~\bibnamefont {Baumert}},\ }\bibfield  {title} {\bibinfo
  {title} {Circular dichroism in the photoelectron angular distributions of
  camphor and fenchone from multiphoton ionization with femtosecond laser
  pulses},\ }\href@noop {} {\bibfield  {journal} {\bibinfo  {journal} {Angew.
  Chem., Int. Ed.}\ }\textbf {\bibinfo {volume} {51}},\ \bibinfo {pages} {5001}
  (\bibinfo {year} {2012})}\BibitemShut {NoStop}%
\bibitem [{\citenamefont {Fischer}\ and\ \citenamefont
  {Hache}(2005)}]{Fischer2005}%
  \BibitemOpen
  \bibfield  {author} {\bibinfo {author} {\bibfnamefont {P.}~\bibnamefont
  {Fischer}}\ and\ \bibinfo {author} {\bibfnamefont {F.}~\bibnamefont
  {Hache}},\ }\bibfield  {title} {\bibinfo {title} {Nonlinear optical
  spectroscopy of chiral molecules},\ }\href@noop {} {\bibfield  {journal}
  {\bibinfo  {journal} {Chirality}\ }\textbf {\bibinfo {volume} {17}},\
  \bibinfo {pages} {421} (\bibinfo {year} {2005})}\BibitemShut {NoStop}%
\bibitem [{\citenamefont {Cireasa}\ \emph {et~al.}(2015)\citenamefont
  {Cireasa}, \citenamefont {Boguslavskiy}, \citenamefont {Pons}, \citenamefont
  {Wong}, \citenamefont {Descamps}, \citenamefont {Petit}, \citenamefont {Ruf},
  \citenamefont {Thir\'e}, \citenamefont {Ferr\'e}, \citenamefont {Suarez},
  \citenamefont {Higuet}, \citenamefont {Schmidt}, \citenamefont {Alharbi},
  \citenamefont {L\'egar\'e}, \citenamefont {Blanchet}, \citenamefont {Fabre},
  \citenamefont {Patchkovskii}, \citenamefont {Smirnova}, \citenamefont
  {Mairesse},\ and\ \citenamefont {Bhardwaj}}]{Cireasa2015}%
  \BibitemOpen
  \bibfield  {author} {\bibinfo {author} {\bibfnamefont {R.}~\bibnamefont
  {Cireasa}}, \bibinfo {author} {\bibfnamefont {A.~E.}\ \bibnamefont
  {Boguslavskiy}}, \bibinfo {author} {\bibfnamefont {B.}~\bibnamefont {Pons}},
  \bibinfo {author} {\bibfnamefont {M.~C.~H.}\ \bibnamefont {Wong}}, \bibinfo
  {author} {\bibfnamefont {D.}~\bibnamefont {Descamps}}, \bibinfo {author}
  {\bibfnamefont {S.}~\bibnamefont {Petit}}, \bibinfo {author} {\bibfnamefont
  {H.}~\bibnamefont {Ruf}}, \bibinfo {author} {\bibfnamefont {N.}~\bibnamefont
  {Thir\'e}}, \bibinfo {author} {\bibfnamefont {A.}~\bibnamefont {Ferr\'e}},
  \bibinfo {author} {\bibfnamefont {J.}~\bibnamefont {Suarez}}, \bibinfo
  {author} {\bibfnamefont {J.}~\bibnamefont {Higuet}}, \bibinfo {author}
  {\bibfnamefont {B.~E.}\ \bibnamefont {Schmidt}}, \bibinfo {author}
  {\bibfnamefont {A.~F.}\ \bibnamefont {Alharbi}}, \bibinfo {author}
  {\bibfnamefont {F.}~\bibnamefont {L\'egar\'e}}, \bibinfo {author}
  {\bibfnamefont {V.}~\bibnamefont {Blanchet}}, \bibinfo {author}
  {\bibfnamefont {B.}~\bibnamefont {Fabre}}, \bibinfo {author} {\bibfnamefont
  {S.}~\bibnamefont {Patchkovskii}}, \bibinfo {author} {\bibfnamefont
  {O.}~\bibnamefont {Smirnova}}, \bibinfo {author} {\bibfnamefont
  {Y.}~\bibnamefont {Mairesse}},\ and\ \bibinfo {author} {\bibfnamefont
  {V.~R.}\ \bibnamefont {Bhardwaj}},\ }\bibfield  {title} {\bibinfo {title}
  {Probing molecular chirality on a subfemtosecond timescale},\ }\href@noop {}
  {\bibfield  {journal} {\bibinfo  {journal} {Nat. Phys.}\ }\textbf {\bibinfo
  {volume} {11}},\ \bibinfo {pages} {654} (\bibinfo {year} {2015})}\BibitemShut
  {NoStop}%
\bibitem [{\citenamefont {Smirnova}\ \emph {et~al.}(2015)\citenamefont
  {Smirnova}, \citenamefont {Mairesse},\ and\ \citenamefont
  {Patchkovskii}}]{Smirnova2015}%
  \BibitemOpen
  \bibfield  {author} {\bibinfo {author} {\bibfnamefont {O.}~\bibnamefont
  {Smirnova}}, \bibinfo {author} {\bibfnamefont {Y.}~\bibnamefont {Mairesse}},\
  and\ \bibinfo {author} {\bibfnamefont {S.}~\bibnamefont {Patchkovskii}},\
  }\bibfield  {title} {\bibinfo {title} {Opportunities for chiral
  discrimination using high harmonic generation in tailored laser fields},\
  }\href@noop {} {\bibfield  {journal} {\bibinfo  {journal} {J. Phys. B}\
  }\textbf {\bibinfo {volume} {48}},\ \bibinfo {pages} {234005} (\bibinfo
  {year} {2015})}\BibitemShut {NoStop}%
\bibitem [{\citenamefont {Baykusheva}\ and\ \citenamefont
  {W\"orner}(2018)}]{Baykusheva2018}%
  \BibitemOpen
  \bibfield  {author} {\bibinfo {author} {\bibfnamefont {D.}~\bibnamefont
  {Baykusheva}}\ and\ \bibinfo {author} {\bibfnamefont {H.~J.}\ \bibnamefont
  {W\"orner}},\ }\bibfield  {title} {\bibinfo {title} {Chiral discrimination
  through bielliptical high-harmonic spectroscopy},\ }\href@noop {} {\bibfield
  {journal} {\bibinfo  {journal} {Phys. Rev. X}\ }\textbf {\bibinfo {volume}
  {8}},\ \bibinfo {pages} {031060} (\bibinfo {year} {2018})}\BibitemShut
  {NoStop}%
\bibitem [{\citenamefont {Neufeld}\ and\ \citenamefont
  {Cohen}(2018)}]{Neufeld2018}%
  \BibitemOpen
  \bibfield  {author} {\bibinfo {author} {\bibfnamefont {O.}~\bibnamefont
  {Neufeld}}\ and\ \bibinfo {author} {\bibfnamefont {O.}~\bibnamefont
  {Cohen}},\ }\bibfield  {title} {\bibinfo {title} {Optical chirality in
  nonlinear optics: Application to high harmonic generation},\ }\href@noop {}
  {\bibfield  {journal} {\bibinfo  {journal} {Phys. Rev. Lett.}\ }\textbf
  {\bibinfo {volume} {120}},\ \bibinfo {pages} {133206} (\bibinfo {year}
  {2018})}\BibitemShut {NoStop}%
\bibitem [{\citenamefont {Harada}\ \emph {et~al.}(2018)\citenamefont {Harada},
  \citenamefont {Haraguchi}, \citenamefont {Kaneshima},\ and\ \citenamefont
  {Sekikawa}}]{Harada2018}%
  \BibitemOpen
  \bibfield  {author} {\bibinfo {author} {\bibfnamefont {Y.}~\bibnamefont
  {Harada}}, \bibinfo {author} {\bibfnamefont {E.}~\bibnamefont {Haraguchi}},
  \bibinfo {author} {\bibfnamefont {K.}~\bibnamefont {Kaneshima}},\ and\
  \bibinfo {author} {\bibfnamefont {T.}~\bibnamefont {Sekikawa}},\ }\bibfield
  {title} {\bibinfo {title} {Circular dichroism in high-order harmonic
  generation from chiral molecules},\ }\href@noop {} {\bibfield  {journal}
  {\bibinfo  {journal} {Phys. Rev. A}\ }\textbf {\bibinfo {volume} {98}},\
  \bibinfo {pages} {021401(R)} (\bibinfo {year} {2018})}\BibitemShut {NoStop}%
\bibitem [{\citenamefont {Neufeld}\ \emph {et~al.}(2019)\citenamefont
  {Neufeld}, \citenamefont {Ayuso}, \citenamefont {Decleva}, \citenamefont
  {Ivanov}, \citenamefont {Smirnova},\ and\ \citenamefont
  {Cohen}}]{Neufeld2019}%
  \BibitemOpen
  \bibfield  {author} {\bibinfo {author} {\bibfnamefont {O.}~\bibnamefont
  {Neufeld}}, \bibinfo {author} {\bibfnamefont {D.}~\bibnamefont {Ayuso}},
  \bibinfo {author} {\bibfnamefont {P.}~\bibnamefont {Decleva}}, \bibinfo
  {author} {\bibfnamefont {M.~Y.}\ \bibnamefont {Ivanov}}, \bibinfo {author}
  {\bibfnamefont {O.}~\bibnamefont {Smirnova}},\ and\ \bibinfo {author}
  {\bibfnamefont {O.}~\bibnamefont {Cohen}},\ }\bibfield  {title} {\bibinfo
  {title} {Ultrasensitive chiral spectroscopy by dynamical symmetry breaking in
  high harmonic generation},\ }\href@noop {} {\bibfield  {journal} {\bibinfo
  {journal} {Phys. Rev. X}\ }\textbf {\bibinfo {volume} {9}},\ \bibinfo {pages}
  {031002} (\bibinfo {year} {2019})}\BibitemShut {NoStop}%
\bibitem [{\citenamefont {Flack}(2003)}]{Flack2003}%
  \BibitemOpen
  \bibfield  {author} {\bibinfo {author} {\bibfnamefont {H.~D.}\ \bibnamefont
  {Flack}},\ }\bibfield  {title} {\bibinfo {title} {Chiral and achiral crystal
  structure},\ }\href@noop {} {\bibfield  {journal} {\bibinfo  {journal} {Helv.
  Chim. Acta}\ }\textbf {\bibinfo {volume} {86}},\ \bibinfo {pages} {905}
  (\bibinfo {year} {2003})}\BibitemShut {NoStop}%
\bibitem [{\citenamefont {Spaldin}\ and\ \citenamefont
  {Fiebig}(2005)}]{Spaldin2005}%
  \BibitemOpen
  \bibfield  {author} {\bibinfo {author} {\bibfnamefont {N.~A.}\ \bibnamefont
  {Spaldin}}\ and\ \bibinfo {author} {\bibfnamefont {M.}~\bibnamefont
  {Fiebig}},\ }\bibfield  {title} {\bibinfo {title} {The renaissance of
  magnetoelectric multiferroics},\ }\href@noop {} {\bibfield  {journal}
  {\bibinfo  {journal} {Science}\ }\textbf {\bibinfo {volume} {309}},\ \bibinfo
  {pages} {391} (\bibinfo {year} {2005})}\BibitemShut {NoStop}%
\bibitem [{\citenamefont {M\"uhlbauer}\ \emph {et~al.}(2009)\citenamefont
  {M\"uhlbauer}, \citenamefont {Binz}, \citenamefont {Jonietz}, \citenamefont
  {Pfleiderer}, \citenamefont {Rosch}, \citenamefont {Neubauer}, \citenamefont
  {Georgii},\ and\ \citenamefont {B\"oni}}]{Muehlbauer2009}%
  \BibitemOpen
  \bibfield  {author} {\bibinfo {author} {\bibfnamefont {S.}~\bibnamefont
  {M\"uhlbauer}}, \bibinfo {author} {\bibfnamefont {B.}~\bibnamefont {Binz}},
  \bibinfo {author} {\bibfnamefont {F.}~\bibnamefont {Jonietz}}, \bibinfo
  {author} {\bibfnamefont {C.}~\bibnamefont {Pfleiderer}}, \bibinfo {author}
  {\bibfnamefont {A.}~\bibnamefont {Rosch}}, \bibinfo {author} {\bibfnamefont
  {A.}~\bibnamefont {Neubauer}}, \bibinfo {author} {\bibfnamefont
  {R.}~\bibnamefont {Georgii}},\ and\ \bibinfo {author} {\bibfnamefont
  {P.}~\bibnamefont {B\"oni}},\ }\bibfield  {title} {\bibinfo {title} {Skyrmion
  lattice in a chiral magnet},\ }\href@noop {} {\bibfield  {journal} {\bibinfo
  {journal} {Science}\ }\textbf {\bibinfo {volume} {323}},\ \bibinfo {pages}
  {915} (\bibinfo {year} {2009})}\BibitemShut {NoStop}%
\bibitem [{\citenamefont {Yoda}\ \emph {et~al.}(2015)\citenamefont {Yoda},
  \citenamefont {Yokoyama},\ and\ \citenamefont {Murakami}}]{Yoda2015}%
  \BibitemOpen
  \bibfield  {author} {\bibinfo {author} {\bibfnamefont {T.}~\bibnamefont
  {Yoda}}, \bibinfo {author} {\bibfnamefont {T.}~\bibnamefont {Yokoyama}},\
  and\ \bibinfo {author} {\bibfnamefont {S.}~\bibnamefont {Murakami}},\
  }\bibfield  {title} {\bibinfo {title} {Current-induced orbital and spin
  magnetizations in crystals with helical structure},\ }\href@noop {}
  {\bibfield  {journal} {\bibinfo  {journal} {Sci. Rep.}\ }\textbf {\bibinfo
  {volume} {5}},\ \bibinfo {pages} {12024} (\bibinfo {year}
  {2015})}\BibitemShut {NoStop}%
\bibitem [{\citenamefont {Tsirkin}\ \emph {et~al.}(2018)\citenamefont
  {Tsirkin}, \citenamefont {Puente},\ and\ \citenamefont
  {Souza}}]{Tsirkin2018}%
  \BibitemOpen
  \bibfield  {author} {\bibinfo {author} {\bibfnamefont {S.~S.}\ \bibnamefont
  {Tsirkin}}, \bibinfo {author} {\bibfnamefont {P.~A.}\ \bibnamefont
  {Puente}},\ and\ \bibinfo {author} {\bibfnamefont {I.}~\bibnamefont
  {Souza}},\ }\bibfield  {title} {\bibinfo {title} {Gyrotropic effects in
  trigonal tellurium studied from first principles},\ }\href@noop {} {\bibfield
   {journal} {\bibinfo  {journal} {Phys. Rev. B}\ }\textbf {\bibinfo {volume}
  {97}},\ \bibinfo {pages} {035158} (\bibinfo {year} {2018})}\BibitemShut
  {NoStop}%
\bibitem [{\citenamefont {Chang}\ \emph {et~al.}(2018)\citenamefont {Chang},
  \citenamefont {Wieder}, \citenamefont {Schindler}, \citenamefont {Sanchez},
  \citenamefont {Belopolski}, \citenamefont {Huang}, \citenamefont {Singh},
  \citenamefont {Wu}, \citenamefont {Chang}, \citenamefont {Neupert},
  \citenamefont {Xu}, \citenamefont {Lin},\ and\ \citenamefont
  {Hasan}}]{Chang2018}%
  \BibitemOpen
  \bibfield  {author} {\bibinfo {author} {\bibfnamefont {G.}~\bibnamefont
  {Chang}}, \bibinfo {author} {\bibfnamefont {B.~J.}\ \bibnamefont {Wieder}},
  \bibinfo {author} {\bibfnamefont {F.}~\bibnamefont {Schindler}}, \bibinfo
  {author} {\bibfnamefont {D.~S.}\ \bibnamefont {Sanchez}}, \bibinfo {author}
  {\bibfnamefont {I.}~\bibnamefont {Belopolski}}, \bibinfo {author}
  {\bibfnamefont {S.-M.}\ \bibnamefont {Huang}}, \bibinfo {author}
  {\bibfnamefont {B.}~\bibnamefont {Singh}}, \bibinfo {author} {\bibfnamefont
  {D.}~\bibnamefont {Wu}}, \bibinfo {author} {\bibfnamefont {T.-R.}\
  \bibnamefont {Chang}}, \bibinfo {author} {\bibfnamefont {T.}~\bibnamefont
  {Neupert}}, \bibinfo {author} {\bibfnamefont {S.-Y.}\ \bibnamefont {Xu}},
  \bibinfo {author} {\bibfnamefont {H.}~\bibnamefont {Lin}},\ and\ \bibinfo
  {author} {\bibfnamefont {M.~Z.}\ \bibnamefont {Hasan}},\ }\bibfield  {title}
  {\bibinfo {title} {Topological quantum properties of chiral crystals},\
  }\href@noop {} {\bibfield  {journal} {\bibinfo  {journal} {Nat. Mater.}\
  }\textbf {\bibinfo {volume} {17}},\ \bibinfo {pages} {978} (\bibinfo {year}
  {2018})}\BibitemShut {NoStop}%
\bibitem [{\citenamefont {Sanchez}\ \emph {et~al.}(2019)\citenamefont
  {Sanchez}, \citenamefont {Belopolski}, \citenamefont {Cochran}, \citenamefont
  {Xu}, \citenamefont {Yin}, \citenamefont {Chang}, \citenamefont {Xie},
  \citenamefont {Manna}, \citenamefont {S\"u{\ss}}, \citenamefont {Huang},
  \citenamefont {Alidoust}, \citenamefont {Multer}, \citenamefont {Zhang},
  \citenamefont {Shumiya}, \citenamefont {Wang}, \citenamefont {Wang},
  \citenamefont {Chang}, \citenamefont {Felser}, \citenamefont {Xu},
  \citenamefont {Jia}, \citenamefont {Lin},\ and\ \citenamefont
  {Hasan}}]{Sanchez2019}%
  \BibitemOpen
  \bibfield  {author} {\bibinfo {author} {\bibfnamefont {D.~S.}\ \bibnamefont
  {Sanchez}}, \bibinfo {author} {\bibfnamefont {I.}~\bibnamefont {Belopolski}},
  \bibinfo {author} {\bibfnamefont {T.~A.}\ \bibnamefont {Cochran}}, \bibinfo
  {author} {\bibfnamefont {X.}~\bibnamefont {Xu}}, \bibinfo {author}
  {\bibfnamefont {J.-X.}\ \bibnamefont {Yin}}, \bibinfo {author} {\bibfnamefont
  {G.}~\bibnamefont {Chang}}, \bibinfo {author} {\bibfnamefont
  {W.}~\bibnamefont {Xie}}, \bibinfo {author} {\bibfnamefont {K.}~\bibnamefont
  {Manna}}, \bibinfo {author} {\bibfnamefont {V.}~\bibnamefont {S\"u{\ss}}},
  \bibinfo {author} {\bibfnamefont {C.-Y.}\ \bibnamefont {Huang}}, \bibinfo
  {author} {\bibfnamefont {N.}~\bibnamefont {Alidoust}}, \bibinfo {author}
  {\bibfnamefont {D.}~\bibnamefont {Multer}}, \bibinfo {author} {\bibfnamefont
  {S.~S.}\ \bibnamefont {Zhang}}, \bibinfo {author} {\bibfnamefont
  {N.}~\bibnamefont {Shumiya}}, \bibinfo {author} {\bibfnamefont
  {X.}~\bibnamefont {Wang}}, \bibinfo {author} {\bibfnamefont {G.-Q.}\
  \bibnamefont {Wang}}, \bibinfo {author} {\bibfnamefont {T.-R.}\ \bibnamefont
  {Chang}}, \bibinfo {author} {\bibfnamefont {C.}~\bibnamefont {Felser}},
  \bibinfo {author} {\bibfnamefont {S.-Y.}\ \bibnamefont {Xu}}, \bibinfo
  {author} {\bibfnamefont {S.}~\bibnamefont {Jia}}, \bibinfo {author}
  {\bibfnamefont {H.}~\bibnamefont {Lin}},\ and\ \bibinfo {author}
  {\bibfnamefont {M.~Z.}\ \bibnamefont {Hasan}},\ }\bibfield  {title} {\bibinfo
  {title} {Topological chiral crystals with helicoid-arc quantum states},\
  }\href@noop {} {\bibfield  {journal} {\bibinfo  {journal} {Nature}\ }\textbf
  {\bibinfo {volume} {567}},\ \bibinfo {pages} {500} (\bibinfo {year}
  {2019})}\BibitemShut {NoStop}%
\bibitem [{\citenamefont {Ghimire}\ \emph {et~al.}(2011)\citenamefont
  {Ghimire}, \citenamefont {DiChiara}, \citenamefont {Sistrunk}, \citenamefont
  {Agostini}, \citenamefont {DiMauro},\ and\ \citenamefont
  {Reis}}]{Ghimire2011}%
  \BibitemOpen
  \bibfield  {author} {\bibinfo {author} {\bibfnamefont {S.}~\bibnamefont
  {Ghimire}}, \bibinfo {author} {\bibfnamefont {A.~D.}\ \bibnamefont
  {DiChiara}}, \bibinfo {author} {\bibfnamefont {E.}~\bibnamefont {Sistrunk}},
  \bibinfo {author} {\bibfnamefont {P.}~\bibnamefont {Agostini}}, \bibinfo
  {author} {\bibfnamefont {L.~F.}\ \bibnamefont {DiMauro}},\ and\ \bibinfo
  {author} {\bibfnamefont {D.~A.}\ \bibnamefont {Reis}},\ }\bibfield  {title}
  {\bibinfo {title} {Observation of high-order harmonic generation in a bulk
  crystal},\ }\href {https://doi.org/10.1038/nphys1847} {\bibfield  {journal}
  {\bibinfo  {journal} {Nat. Phys.}\ }\textbf {\bibinfo {volume} {7}},\
  \bibinfo {pages} {138} (\bibinfo {year} {2011})}\BibitemShut {NoStop}%
\bibitem [{\citenamefont {Ghimire}\ and\ \citenamefont
  {Reis}(2019)}]{Ghimire2019}%
  \BibitemOpen
  \bibfield  {author} {\bibinfo {author} {\bibfnamefont {S.}~\bibnamefont
  {Ghimire}}\ and\ \bibinfo {author} {\bibfnamefont {D.~A.}\ \bibnamefont
  {Reis}},\ }\bibfield  {title} {\bibinfo {title} {High-harmonic generation
  from solids},\ }\href@noop {} {\bibfield  {journal} {\bibinfo  {journal}
  {Nat. Phys.}\ }\textbf {\bibinfo {volume} {15}},\ \bibinfo {pages} {10}
  (\bibinfo {year} {2019})}\BibitemShut {NoStop}%
\bibitem [{\citenamefont {Tancogne-Dejean}\ \emph
  {et~al.}(2017{\natexlab{a}})\citenamefont {Tancogne-Dejean}, \citenamefont
  {M\"ucke}, \citenamefont {K\"artner},\ and\ \citenamefont {Rubio}}]{TD2017b}%
  \BibitemOpen
  \bibfield  {author} {\bibinfo {author} {\bibfnamefont {N.}~\bibnamefont
  {Tancogne-Dejean}}, \bibinfo {author} {\bibfnamefont {O.~D.}\ \bibnamefont
  {M\"ucke}}, \bibinfo {author} {\bibfnamefont {F.~X.}\ \bibnamefont
  {K\"artner}},\ and\ \bibinfo {author} {\bibfnamefont {A.}~\bibnamefont
  {Rubio}},\ }\bibfield  {title} {\bibinfo {title} {Ellipticity dependence of
  high-harmonic generation in solids originating from coupled intraband and
  interband dynamics},\ }\href {https://doi.org/10.1038/s41467-017-00764-5}
  {\bibfield  {journal} {\bibinfo  {journal} {Nat. Commun.}\ }\textbf {\bibinfo
  {volume} {8}},\ \bibinfo {pages} {745} (\bibinfo {year}
  {2017}{\natexlab{a}})}\BibitemShut {NoStop}%
\bibitem [{\citenamefont {Langer}\ \emph {et~al.}(2017)\citenamefont {Langer},
  \citenamefont {Hohenleutner}, \citenamefont {Huttner}, \citenamefont {Koch},
  \citenamefont {Kira},\ and\ \citenamefont {Huber}}]{Langer2017}%
  \BibitemOpen
  \bibfield  {author} {\bibinfo {author} {\bibfnamefont {F.}~\bibnamefont
  {Langer}}, \bibinfo {author} {\bibfnamefont {M.}~\bibnamefont
  {Hohenleutner}}, \bibinfo {author} {\bibfnamefont {U.}~\bibnamefont
  {Huttner}}, \bibinfo {author} {\bibfnamefont {S.~W.}\ \bibnamefont {Koch}},
  \bibinfo {author} {\bibfnamefont {M.}~\bibnamefont {Kira}},\ and\ \bibinfo
  {author} {\bibfnamefont {R.}~\bibnamefont {Huber}},\ }\bibfield  {title}
  {\bibinfo {title} {Symmetry-controlled temporal structure of high-harmonic
  carrier fields from a bulk crystal},\ }\href
  {https://doi.org/10.1038/nphoton.2017.29} {\bibfield  {journal} {\bibinfo
  {journal} {Nat. Photonics}\ }\textbf {\bibinfo {volume} {11}},\ \bibinfo
  {pages} {227} (\bibinfo {year} {2017})}\BibitemShut {NoStop}%
\bibitem [{\citenamefont {Hammond}\ \emph {et~al.}(2017)\citenamefont
  {Hammond}, \citenamefont {Monchoc\"e}, \citenamefont {Zhang}, \citenamefont
  {Vampa}, \citenamefont {Klug}, \citenamefont {Naumov}, \citenamefont
  {Villeneuve},\ and\ \citenamefont {Corkum}}]{Hammond2017}%
  \BibitemOpen
  \bibfield  {author} {\bibinfo {author} {\bibfnamefont {T.~J.}\ \bibnamefont
  {Hammond}}, \bibinfo {author} {\bibfnamefont {S.}~\bibnamefont {Monchoc\"e}},
  \bibinfo {author} {\bibfnamefont {C.}~\bibnamefont {Zhang}}, \bibinfo
  {author} {\bibfnamefont {G.}~\bibnamefont {Vampa}}, \bibinfo {author}
  {\bibfnamefont {D.}~\bibnamefont {Klug}}, \bibinfo {author} {\bibfnamefont
  {A.~Y.}\ \bibnamefont {Naumov}}, \bibinfo {author} {\bibfnamefont {D.~M.}\
  \bibnamefont {Villeneuve}},\ and\ \bibinfo {author} {\bibfnamefont {P.~B.}\
  \bibnamefont {Corkum}},\ }\bibfield  {title} {\bibinfo {title} {Integrating
  solids and gases for attosecond pulse generation},\ }\href
  {https://doi.org/10.1038/nphoton.2017.141} {\bibfield  {journal} {\bibinfo
  {journal} {Nat. Photonics}\ }\textbf {\bibinfo {volume} {11}},\ \bibinfo
  {pages} {594} (\bibinfo {year} {2017})}\BibitemShut {NoStop}%
\bibitem [{\citenamefont {Sivis}\ \emph {et~al.}(2017)\citenamefont {Sivis},
  \citenamefont {Taucer}, \citenamefont {Vampa}, \citenamefont {Johnston},
  \citenamefont {Staudte}, \citenamefont {Naumov}, \citenamefont {Villeneuve},
  \citenamefont {Ropers},\ and\ \citenamefont {Corkum}}]{Sivis2017}%
  \BibitemOpen
  \bibfield  {author} {\bibinfo {author} {\bibfnamefont {M.}~\bibnamefont
  {Sivis}}, \bibinfo {author} {\bibfnamefont {M.}~\bibnamefont {Taucer}},
  \bibinfo {author} {\bibfnamefont {G.}~\bibnamefont {Vampa}}, \bibinfo
  {author} {\bibfnamefont {K.}~\bibnamefont {Johnston}}, \bibinfo {author}
  {\bibfnamefont {A.}~\bibnamefont {Staudte}}, \bibinfo {author} {\bibfnamefont
  {A.~Y.}\ \bibnamefont {Naumov}}, \bibinfo {author} {\bibfnamefont {D.~M.}\
  \bibnamefont {Villeneuve}}, \bibinfo {author} {\bibfnamefont
  {C.}~\bibnamefont {Ropers}},\ and\ \bibinfo {author} {\bibfnamefont {P.~B.}\
  \bibnamefont {Corkum}},\ }\bibfield  {title} {\bibinfo {title} {Tailored
  semiconductors for high-harmonic optoelectronics},\ }\href
  {https://doi.org/10.1126/science.aan2395} {\bibfield  {journal} {\bibinfo
  {journal} {Science}\ }\textbf {\bibinfo {volume} {357}},\ \bibinfo {pages}
  {303} (\bibinfo {year} {2017})}\BibitemShut {NoStop}%
\bibitem [{\citenamefont {Garg}\ \emph {et~al.}(2018)\citenamefont {Garg},
  \citenamefont {Kim},\ and\ \citenamefont {Goulielmakis}}]{Garg2018}%
  \BibitemOpen
  \bibfield  {author} {\bibinfo {author} {\bibfnamefont {M.}~\bibnamefont
  {Garg}}, \bibinfo {author} {\bibfnamefont {H.~Y.}\ \bibnamefont {Kim}},\ and\
  \bibinfo {author} {\bibfnamefont {E.}~\bibnamefont {Goulielmakis}},\
  }\bibfield  {title} {\bibinfo {title} {Ultimate waveform reproducibility of
  extreme-ultraviolet pulses by high-harmonic generation in quartz},\ }\href
  {https://doi.org/10.1038/s41566-018-0123-6} {\bibfield  {journal} {\bibinfo
  {journal} {Nat. Photonics}\ }\textbf {\bibinfo {volume} {12}},\ \bibinfo
  {pages} {291} (\bibinfo {year} {2018})}\BibitemShut {NoStop}%
\bibitem [{\citenamefont {Yoshikawa}\ \emph {et~al.}(2017)\citenamefont
  {Yoshikawa}, \citenamefont {Tamaya},\ and\ \citenamefont
  {Tanaka}}]{Yoshikawa2017}%
  \BibitemOpen
  \bibfield  {author} {\bibinfo {author} {\bibfnamefont {N.}~\bibnamefont
  {Yoshikawa}}, \bibinfo {author} {\bibfnamefont {T.}~\bibnamefont {Tamaya}},\
  and\ \bibinfo {author} {\bibfnamefont {K.}~\bibnamefont {Tanaka}},\
  }\bibfield  {title} {\bibinfo {title} {High-harmonic generation in graphene
  enhanced by elliptically polarized light excitation},\ }\href
  {https://doi.org/10.1126/science.aam8861} {\bibfield  {journal} {\bibinfo
  {journal} {Science}\ }\textbf {\bibinfo {volume} {356}},\ \bibinfo {pages}
  {736} (\bibinfo {year} {2017})}\BibitemShut {NoStop}%
\bibitem [{\citenamefont {Taucer}\ \emph {et~al.}(2017)\citenamefont {Taucer},
  \citenamefont {Hammond}, \citenamefont {Corkum}, \citenamefont {Vampa},
  \citenamefont {Couture}, \citenamefont {Thir\'e}, \citenamefont {Schmidt},
  \citenamefont {L\'egar\'e}, \citenamefont {Selvi}, \citenamefont {Unsuree},
  \citenamefont {Hamilton}, \citenamefont {Echtermeyer},\ and\ \citenamefont
  {Denecke}}]{Taucer2017}%
  \BibitemOpen
  \bibfield  {author} {\bibinfo {author} {\bibfnamefont {M.}~\bibnamefont
  {Taucer}}, \bibinfo {author} {\bibfnamefont {T.~J.}\ \bibnamefont {Hammond}},
  \bibinfo {author} {\bibfnamefont {P.~B.}\ \bibnamefont {Corkum}}, \bibinfo
  {author} {\bibfnamefont {G.}~\bibnamefont {Vampa}}, \bibinfo {author}
  {\bibfnamefont {C.}~\bibnamefont {Couture}}, \bibinfo {author} {\bibfnamefont
  {N.}~\bibnamefont {Thir\'e}}, \bibinfo {author} {\bibfnamefont {B.~E.}\
  \bibnamefont {Schmidt}}, \bibinfo {author} {\bibfnamefont {F.}~\bibnamefont
  {L\'egar\'e}}, \bibinfo {author} {\bibfnamefont {H.}~\bibnamefont {Selvi}},
  \bibinfo {author} {\bibfnamefont {N.}~\bibnamefont {Unsuree}}, \bibinfo
  {author} {\bibfnamefont {B.}~\bibnamefont {Hamilton}}, \bibinfo {author}
  {\bibfnamefont {T.~J.}\ \bibnamefont {Echtermeyer}},\ and\ \bibinfo {author}
  {\bibfnamefont {M.~A.}\ \bibnamefont {Denecke}},\ }\bibfield  {title}
  {\bibinfo {title} {Nonperturbative harmonic generation in graphene from
  intense midinfrared pulsed light},\ }\href
  {https://doi.org/10.1103/physrevb.96.195420} {\bibfield  {journal} {\bibinfo
  {journal} {Phys. Rev. B}\ }\textbf {\bibinfo {volume} {96}},\ \bibinfo
  {pages} {195420} (\bibinfo {year} {2017})}\BibitemShut {NoStop}%
\bibitem [{\citenamefont {Baudisch}\ \emph {et~al.}(2018)\citenamefont
  {Baudisch}, \citenamefont {Marini}, \citenamefont {Cox}, \citenamefont {Zhu},
  \citenamefont {Silva}, \citenamefont {Teichmann}, \citenamefont {Massicotte},
  \citenamefont {Koppens}, \citenamefont {Levitov}, \citenamefont {Garc\'ia~de
  Abajo},\ and\ \citenamefont {Jens}}]{Baudisch2018}%
  \BibitemOpen
  \bibfield  {author} {\bibinfo {author} {\bibfnamefont {M.}~\bibnamefont
  {Baudisch}}, \bibinfo {author} {\bibfnamefont {A.}~\bibnamefont {Marini}},
  \bibinfo {author} {\bibfnamefont {J.~D.}\ \bibnamefont {Cox}}, \bibinfo
  {author} {\bibfnamefont {T.}~\bibnamefont {Zhu}}, \bibinfo {author}
  {\bibfnamefont {F.}~\bibnamefont {Silva}}, \bibinfo {author} {\bibfnamefont
  {S.}~\bibnamefont {Teichmann}}, \bibinfo {author} {\bibfnamefont
  {M.}~\bibnamefont {Massicotte}}, \bibinfo {author} {\bibfnamefont
  {F.}~\bibnamefont {Koppens}}, \bibinfo {author} {\bibfnamefont {L.~S.}\
  \bibnamefont {Levitov}}, \bibinfo {author} {\bibfnamefont {F.~J.}\
  \bibnamefont {Garc\'ia~de Abajo}},\ and\ \bibinfo {author} {\bibfnamefont
  {B.}~\bibnamefont {Jens}},\ }\bibfield  {title} {\bibinfo {title} {Ultrafast
  nonlinear optical response of dirac fermions in graphene},\ }\href
  {https://doi.org/10.1038/s41467-018-03413-7} {\bibfield  {journal} {\bibinfo
  {journal} {Nat. Commun.}\ }\textbf {\bibinfo {volume} {9}},\ \bibinfo {pages}
  {1018} (\bibinfo {year} {2018})}\BibitemShut {NoStop}%
\bibitem [{\citenamefont {Chen}\ and\ \citenamefont
  {Qin}(2019{\natexlab{a}})}]{chen_circularly_2019}%
  \BibitemOpen
  \bibfield  {author} {\bibinfo {author} {\bibfnamefont {Z.-Y.}\ \bibnamefont
  {Chen}}\ and\ \bibinfo {author} {\bibfnamefont {R.}~\bibnamefont {Qin}},\
  }\bibfield  {title} {\bibinfo {title} {Circularly polarized extreme
  ultraviolet high harmonic generation in graphene},\ }\href
  {https://doi.org/10.1364/OE.27.003761} {\bibfield  {journal} {\bibinfo
  {journal} {Opt. Express}\ }\textbf {\bibinfo {volume} {27}},\ \bibinfo
  {pages} {3761} (\bibinfo {year} {2019}{\natexlab{a}})}\BibitemShut {NoStop}%
\bibitem [{\citenamefont {Qin}\ and\ \citenamefont {Chen}(2018)}]{Qin2018}%
  \BibitemOpen
  \bibfield  {author} {\bibinfo {author} {\bibfnamefont {R.}~\bibnamefont
  {Qin}}\ and\ \bibinfo {author} {\bibfnamefont {Z.-Y.}\ \bibnamefont {Chen}},\
  }\bibfield  {title} {\bibinfo {title} {Strain-controlled high harmonic
  generation with dirac fermions in silicene},\ }\href {https://doi.org/DOI:
  10.1039/C8NR07572G} {\bibfield  {journal} {\bibinfo  {journal} {Nanoscale}\
  }\textbf {\bibinfo {volume} {10}},\ \bibinfo {pages} {22593} (\bibinfo {year}
  {2018})}\BibitemShut {NoStop}%
\bibitem [{\citenamefont {Chen}\ and\ \citenamefont
  {Qin}(2019{\natexlab{b}})}]{Chen_BP_2019}%
  \BibitemOpen
  \bibfield  {author} {\bibinfo {author} {\bibfnamefont {Z.-Y.}\ \bibnamefont
  {Chen}}\ and\ \bibinfo {author} {\bibfnamefont {R.}~\bibnamefont {Qin}},\
  }\bibfield  {title} {\bibinfo {title} {Strong-field nonlinear optical
  properties of monolayer black phosphorus},\ }\href@noop {} {\bibfield
  {journal} {\bibinfo  {journal} {Nanoscale}\ }\textbf {\bibinfo {volume}
  {11}},\ \bibinfo {pages} {16377} (\bibinfo {year}
  {2019}{\natexlab{b}})}\BibitemShut {NoStop}%
\bibitem [{\citenamefont {Guan}\ \emph {et~al.}(2019)\citenamefont {Guan},
  \citenamefont {Lian}, \citenamefont {Hu}, \citenamefont {Liu}, \citenamefont
  {Zhang}, \citenamefont {Zhang},\ and\ \citenamefont
  {Meng}}]{guan_cooperative_2019}%
  \BibitemOpen
  \bibfield  {author} {\bibinfo {author} {\bibfnamefont {M.-X.}\ \bibnamefont
  {Guan}}, \bibinfo {author} {\bibfnamefont {C.}~\bibnamefont {Lian}}, \bibinfo
  {author} {\bibfnamefont {S.-Q.}\ \bibnamefont {Hu}}, \bibinfo {author}
  {\bibfnamefont {H.}~\bibnamefont {Liu}}, \bibinfo {author} {\bibfnamefont
  {S.-J.}\ \bibnamefont {Zhang}}, \bibinfo {author} {\bibfnamefont
  {J.}~\bibnamefont {Zhang}},\ and\ \bibinfo {author} {\bibfnamefont
  {S.}~\bibnamefont {Meng}},\ }\bibfield  {title} {\bibinfo {title}
  {Cooperative evolution of intraband and interband excitations for
  high-harmonic generation in strained {MoS2}},\ }\href@noop {} {\bibfield
  {journal} {\bibinfo  {journal} {Phys. Rev. B}\ }\textbf {\bibinfo {volume}
  {99}},\ \bibinfo {pages} {184306} (\bibinfo {year} {2019})}\BibitemShut
  {NoStop}%
\bibitem [{\citenamefont {Guan}\ \emph {et~al.}(2020)\citenamefont {Guan},
  \citenamefont {Hu}, \citenamefont {Zhao}, \citenamefont {Lian},\ and\
  \citenamefont {Meng}}]{guan_toward_2020}%
  \BibitemOpen
  \bibfield  {author} {\bibinfo {author} {\bibfnamefont {M.}~\bibnamefont
  {Guan}}, \bibinfo {author} {\bibfnamefont {S.}~\bibnamefont {Hu}}, \bibinfo
  {author} {\bibfnamefont {H.}~\bibnamefont {Zhao}}, \bibinfo {author}
  {\bibfnamefont {C.}~\bibnamefont {Lian}},\ and\ \bibinfo {author}
  {\bibfnamefont {S.}~\bibnamefont {Meng}},\ }\bibfield  {title} {\bibinfo
  {title} {Toward attosecond control of electron dynamics in two-dimensional
  materials},\ }\href@noop {} {\bibfield  {journal} {\bibinfo  {journal} {Appl.
  Phys. Lett.}\ }\textbf {\bibinfo {volume} {116}},\ \bibinfo {pages} {043101}
  (\bibinfo {year} {2020})}\BibitemShut {NoStop}%
\bibitem [{\citenamefont {Yoshikawa}\ \emph {et~al.}(2019)\citenamefont
  {Yoshikawa}, \citenamefont {Nagai}, \citenamefont {Uchida}, \citenamefont
  {Takaguchi}, \citenamefont {Sasaki}, \citenamefont {Miyata},\ and\
  \citenamefont {Tanaka}}]{yoshikawa_interband_2019}%
  \BibitemOpen
  \bibfield  {author} {\bibinfo {author} {\bibfnamefont {N.}~\bibnamefont
  {Yoshikawa}}, \bibinfo {author} {\bibfnamefont {K.}~\bibnamefont {Nagai}},
  \bibinfo {author} {\bibfnamefont {K.}~\bibnamefont {Uchida}}, \bibinfo
  {author} {\bibfnamefont {Y.}~\bibnamefont {Takaguchi}}, \bibinfo {author}
  {\bibfnamefont {S.}~\bibnamefont {Sasaki}}, \bibinfo {author} {\bibfnamefont
  {Y.}~\bibnamefont {Miyata}},\ and\ \bibinfo {author} {\bibfnamefont
  {K.}~\bibnamefont {Tanaka}},\ }\bibfield  {title} {\bibinfo {title}
  {Interband resonant high-harmonic generation by valley polarized
  electron–hole pairs},\ }\href@noop {} {\bibfield  {journal} {\bibinfo
  {journal} {Nat. Commun.}\ }\textbf {\bibinfo {volume} {10}},\ \bibinfo
  {pages} {3709} (\bibinfo {year} {2019})}\BibitemShut {NoStop}%
\bibitem [{\citenamefont {Tancogne-Dejean}\ and\ \citenamefont
  {Rubio}(2018)}]{TD2018}%
  \BibitemOpen
  \bibfield  {author} {\bibinfo {author} {\bibfnamefont {N.}~\bibnamefont
  {Tancogne-Dejean}}\ and\ \bibinfo {author} {\bibfnamefont {A.}~\bibnamefont
  {Rubio}},\ }\bibfield  {title} {\bibinfo {title} {Atomic-like high-harmonic
  generation from two-dimensional materials},\ }\href
  {https://doi.org/10.1126/sciadv.aao5207} {\bibfield  {journal} {\bibinfo
  {journal} {Sci. Adv.}\ }\textbf {\bibinfo {volume} {4}},\ \bibinfo {pages}
  {eaao5207} (\bibinfo {year} {2018})}\BibitemShut {NoStop}%
\bibitem [{\citenamefont {Le~Breton}\ \emph {et~al.}(2018)\citenamefont
  {Le~Breton}, \citenamefont {Rubio},\ and\ \citenamefont
  {Tancogne-Dejean}}]{le_breton_high-harmonic_2018}%
  \BibitemOpen
  \bibfield  {author} {\bibinfo {author} {\bibfnamefont {G.}~\bibnamefont
  {Le~Breton}}, \bibinfo {author} {\bibfnamefont {A.}~\bibnamefont {Rubio}},\
  and\ \bibinfo {author} {\bibfnamefont {N.}~\bibnamefont {Tancogne-Dejean}},\
  }\bibfield  {title} {\bibinfo {title} {High-harmonic generation from
  few-layer hexagonal boron nitride: {Evolution} from monolayer to bulk
  response},\ }\href {https://doi.org/10.1103/PhysRevB.98.165308} {\bibfield
  {journal} {\bibinfo  {journal} {Phys. Rev. B}\ }\textbf {\bibinfo {volume}
  {98}},\ \bibinfo {pages} {165308} (\bibinfo {year} {2018})}\BibitemShut
  {NoStop}%
\bibitem [{\citenamefont {Vampa}\ \emph {et~al.}(2014)\citenamefont {Vampa},
  \citenamefont {McDonald}, \citenamefont {Orlando}, \citenamefont {Klug},
  \citenamefont {Corkum},\ and\ \citenamefont {Brabec}}]{Vampa2014}%
  \BibitemOpen
  \bibfield  {author} {\bibinfo {author} {\bibfnamefont {G.}~\bibnamefont
  {Vampa}}, \bibinfo {author} {\bibfnamefont {C.~R.}\ \bibnamefont {McDonald}},
  \bibinfo {author} {\bibfnamefont {G.}~\bibnamefont {Orlando}}, \bibinfo
  {author} {\bibfnamefont {D.~D.}\ \bibnamefont {Klug}}, \bibinfo {author}
  {\bibfnamefont {P.~B.}\ \bibnamefont {Corkum}},\ and\ \bibinfo {author}
  {\bibfnamefont {T.}~\bibnamefont {Brabec}},\ }\bibfield  {title} {\bibinfo
  {title} {Theoretical analysis of high-harmonic generation in solids},\
  }\href@noop {} {\bibfield  {journal} {\bibinfo  {journal} {Phys. Rev. Lett.}\
  }\textbf {\bibinfo {volume} {113}},\ \bibinfo {pages} {073901} (\bibinfo
  {year} {2014})}\BibitemShut {NoStop}%
\bibitem [{\citenamefont {Luu}\ \emph {et~al.}(2015)\citenamefont {Luu},
  \citenamefont {Garg}, \citenamefont {Kruchinin}, \citenamefont {Moulet},
  \citenamefont {Hassan},\ and\ \citenamefont {Goulielmakis}}]{Luu2015}%
  \BibitemOpen
  \bibfield  {author} {\bibinfo {author} {\bibfnamefont {T.~T.}\ \bibnamefont
  {Luu}}, \bibinfo {author} {\bibfnamefont {M.}~\bibnamefont {Garg}}, \bibinfo
  {author} {\bibfnamefont {S.~Y.}\ \bibnamefont {Kruchinin}}, \bibinfo {author}
  {\bibfnamefont {A.}~\bibnamefont {Moulet}}, \bibinfo {author} {\bibfnamefont
  {M.~T.}\ \bibnamefont {Hassan}},\ and\ \bibinfo {author} {\bibfnamefont
  {E.}~\bibnamefont {Goulielmakis}},\ }\bibfield  {title} {\bibinfo {title}
  {Extreme ultraviolet high-harmonic spectroscopy of solids},\ }\href
  {https://doi.org/10.1038/nature14456} {\bibfield  {journal} {\bibinfo
  {journal} {Nature}\ }\textbf {\bibinfo {volume} {521}},\ \bibinfo {pages}
  {498} (\bibinfo {year} {2015})}\BibitemShut {NoStop}%
\bibitem [{\citenamefont {Tancogne-Dejean}\ \emph
  {et~al.}(2017{\natexlab{b}})\citenamefont {Tancogne-Dejean}, \citenamefont
  {M\"ucke}, \citenamefont {K\"artner},\ and\ \citenamefont {Rubio}}]{TD2017a}%
  \BibitemOpen
  \bibfield  {author} {\bibinfo {author} {\bibfnamefont {N.}~\bibnamefont
  {Tancogne-Dejean}}, \bibinfo {author} {\bibfnamefont {O.~D.}\ \bibnamefont
  {M\"ucke}}, \bibinfo {author} {\bibfnamefont {F.~X.}\ \bibnamefont
  {K\"artner}},\ and\ \bibinfo {author} {\bibfnamefont {A.}~\bibnamefont
  {Rubio}},\ }\bibfield  {title} {\bibinfo {title} {Impact of the electronic
  band structure in high-harmonic generation spectra of solids},\ }\href
  {https://doi.org/10.1103/physrevlett.118.087403} {\bibfield  {journal}
  {\bibinfo  {journal} {Phys. Rev. Lett.}\ }\textbf {\bibinfo {volume} {118}},\
  \bibinfo {pages} {087403} (\bibinfo {year} {2017}{\natexlab{b}})}\BibitemShut
  {NoStop}%
\bibitem [{\citenamefont {Klemke}\ \emph {et~al.}(2019)\citenamefont {Klemke},
  \citenamefont {Tancogne-Dejean}, \citenamefont {Rossi}, \citenamefont {Yang},
  \citenamefont {Scheiba}, \citenamefont {Mainz}, \citenamefont {Di~Sciacca},
  \citenamefont {Rubio}, \citenamefont {Kärtner},\ and\ \citenamefont
  {Mücke}}]{klemke_polarization-state-resolved_2019}%
  \BibitemOpen
  \bibfield  {author} {\bibinfo {author} {\bibfnamefont {N.}~\bibnamefont
  {Klemke}}, \bibinfo {author} {\bibfnamefont {N.}~\bibnamefont
  {Tancogne-Dejean}}, \bibinfo {author} {\bibfnamefont {G.~M.}\ \bibnamefont
  {Rossi}}, \bibinfo {author} {\bibfnamefont {Y.}~\bibnamefont {Yang}},
  \bibinfo {author} {\bibfnamefont {F.}~\bibnamefont {Scheiba}}, \bibinfo
  {author} {\bibfnamefont {R.~E.}\ \bibnamefont {Mainz}}, \bibinfo {author}
  {\bibfnamefont {G.}~\bibnamefont {Di~Sciacca}}, \bibinfo {author}
  {\bibfnamefont {A.}~\bibnamefont {Rubio}}, \bibinfo {author} {\bibfnamefont
  {F.~X.}\ \bibnamefont {Kärtner}},\ and\ \bibinfo {author} {\bibfnamefont
  {O.~D.}\ \bibnamefont {Mücke}},\ }\bibfield  {title} {\bibinfo {title}
  {Polarization-state-resolved high-harmonic spectroscopy of solids},\ }\href
  {https://doi.org/10.1038/s41467-019-09328-1} {\bibfield  {journal} {\bibinfo
  {journal} {Nat. Commun.}\ }\textbf {\bibinfo {volume} {10}},\ \bibinfo
  {pages} {1319} (\bibinfo {year} {2019})}\BibitemShut {NoStop}%
\bibitem [{\citenamefont {You}\ \emph {et~al.}(2017)\citenamefont {You},
  \citenamefont {Reis},\ and\ \citenamefont {Ghimire}}]{You2017}%
  \BibitemOpen
  \bibfield  {author} {\bibinfo {author} {\bibfnamefont {Y.~S.}\ \bibnamefont
  {You}}, \bibinfo {author} {\bibfnamefont {D.~A.}\ \bibnamefont {Reis}},\ and\
  \bibinfo {author} {\bibfnamefont {S.}~\bibnamefont {Ghimire}},\ }\bibfield
  {title} {\bibinfo {title} {Anisotropic high-harmonic generation in bulk
  crystals},\ }\href {https://doi.org/10.1038/nphys3955} {\bibfield  {journal}
  {\bibinfo  {journal} {Nat. Phys.}\ }\textbf {\bibinfo {volume} {13}},\
  \bibinfo {pages} {345} (\bibinfo {year} {2017})}\BibitemShut {NoStop}%
\bibitem [{\citenamefont {Vampa}\ \emph {et~al.}(2015)\citenamefont {Vampa},
  \citenamefont {Hammond}, \citenamefont {Thir{\'{e}}}, \citenamefont
  {Schmidt}, \citenamefont {L{\'{e}}gar{\'{e}}}, \citenamefont {McDonald},
  \citenamefont {Brabec}, \citenamefont {Klug},\ and\ \citenamefont
  {Corkum}}]{Vampa2015b}%
  \BibitemOpen
  \bibfield  {author} {\bibinfo {author} {\bibfnamefont {G.}~\bibnamefont
  {Vampa}}, \bibinfo {author} {\bibfnamefont {T.}~\bibnamefont {Hammond}},
  \bibinfo {author} {\bibfnamefont {N.}~\bibnamefont {Thir{\'{e}}}}, \bibinfo
  {author} {\bibfnamefont {B.}~\bibnamefont {Schmidt}}, \bibinfo {author}
  {\bibfnamefont {F.}~\bibnamefont {L{\'{e}}gar{\'{e}}}}, \bibinfo {author}
  {\bibfnamefont {C.}~\bibnamefont {McDonald}}, \bibinfo {author}
  {\bibfnamefont {T.}~\bibnamefont {Brabec}}, \bibinfo {author} {\bibfnamefont
  {D.}~\bibnamefont {Klug}},\ and\ \bibinfo {author} {\bibfnamefont
  {P.}~\bibnamefont {Corkum}},\ }\bibfield  {title} {\bibinfo {title}
  {All-optical reconstruction of crystal band structure},\ }\href
  {https://doi.org/10.1103/physrevlett.115.193603} {\bibfield  {journal}
  {\bibinfo  {journal} {Phys. Rev. Lett.}\ }\textbf {\bibinfo {volume} {115}},\
  \bibinfo {pages} {193603} (\bibinfo {year} {2015})}\BibitemShut {NoStop}%
\bibitem [{\citenamefont {Lanin}\ \emph {et~al.}(2017)\citenamefont {Lanin},
  \citenamefont {Stepanov}, \citenamefont {Fedotov},\ and\ \citenamefont
  {Zheltikov}}]{lanin_mapping_2017}%
  \BibitemOpen
  \bibfield  {author} {\bibinfo {author} {\bibfnamefont {A.~A.}\ \bibnamefont
  {Lanin}}, \bibinfo {author} {\bibfnamefont {E.~A.}\ \bibnamefont {Stepanov}},
  \bibinfo {author} {\bibfnamefont {A.~B.}\ \bibnamefont {Fedotov}},\ and\
  \bibinfo {author} {\bibfnamefont {A.~M.}\ \bibnamefont {Zheltikov}},\
  }\bibfield  {title} {\bibinfo {title} {Mapping the electron band structure by
  intraband high-harmonic generation in solids},\ }\href
  {https://doi.org/10.1364/OPTICA.4.000516} {\bibfield  {journal} {\bibinfo
  {journal} {Optica}\ }\textbf {\bibinfo {volume} {4}},\ \bibinfo {pages} {516}
  (\bibinfo {year} {2017})}\BibitemShut {NoStop}%
\bibitem [{\citenamefont {Luu}\ and\ \citenamefont {W\"orner}(2018)}]{Luu2018}%
  \BibitemOpen
  \bibfield  {author} {\bibinfo {author} {\bibfnamefont {T.~T.}\ \bibnamefont
  {Luu}}\ and\ \bibinfo {author} {\bibfnamefont {H.~J.}\ \bibnamefont
  {W\"orner}},\ }\bibfield  {title} {\bibinfo {title} {Measurement of the berry
  curvature of solids using high-harmonic spectroscopy},\ }\href
  {https://doi.org/10.1038/s41467-018-03397-4} {\bibfield  {journal} {\bibinfo
  {journal} {Nat. Commun.}\ }\textbf {\bibinfo {volume} {9}},\ \bibinfo {pages}
  {916} (\bibinfo {year} {2018})}\BibitemShut {NoStop}%
\bibitem [{\citenamefont {Silva}\ \emph {et~al.}(2018)\citenamefont {Silva},
  \citenamefont {Blinov}, \citenamefont {Rubtsov}, \citenamefont {Smirnova},\
  and\ \citenamefont {Ivanov}}]{silva_high-harmonic_2018}%
  \BibitemOpen
  \bibfield  {author} {\bibinfo {author} {\bibfnamefont {R.~E.~F.}\
  \bibnamefont {Silva}}, \bibinfo {author} {\bibfnamefont {I.~V.}\ \bibnamefont
  {Blinov}}, \bibinfo {author} {\bibfnamefont {A.~N.}\ \bibnamefont {Rubtsov}},
  \bibinfo {author} {\bibfnamefont {O.}~\bibnamefont {Smirnova}},\ and\
  \bibinfo {author} {\bibfnamefont {M.}~\bibnamefont {Ivanov}},\ }\bibfield
  {title} {\bibinfo {title} {High-harmonic spectroscopy of ultrafast many-body
  dynamics in strongly correlated systems},\ }\href
  {https://doi.org/10.1038/s41566-018-0129-0} {\bibfield  {journal} {\bibinfo
  {journal} {Nat. Photonics}\ }\textbf {\bibinfo {volume} {12}},\ \bibinfo
  {pages} {266} (\bibinfo {year} {2018})}\BibitemShut {NoStop}%
\bibitem [{\citenamefont {Bauer}\ and\ \citenamefont
  {Hansen}(2018)}]{bauer_high-harmonic_2018}%
  \BibitemOpen
  \bibfield  {author} {\bibinfo {author} {\bibfnamefont {D.}~\bibnamefont
  {Bauer}}\ and\ \bibinfo {author} {\bibfnamefont {K.~K.}\ \bibnamefont
  {Hansen}},\ }\bibfield  {title} {\bibinfo {title} {High-{Harmonic}
  {Generation} in {Solids} with and without {Topological} {Edge} {States}},\
  }\href@noop {} {\bibfield  {journal} {\bibinfo  {journal} {Phys. Rev. Lett.}\
  }\textbf {\bibinfo {volume} {120}},\ \bibinfo {pages} {177401} (\bibinfo
  {year} {2018})}\BibitemShut {NoStop}%
\bibitem [{\citenamefont {Silva}\ \emph {et~al.}(2019)\citenamefont {Silva},
  \citenamefont {Jim\'{e}nez-Gal\'{a}n}, \citenamefont {Amorim}, \citenamefont
  {Smirnova},\ and\ \citenamefont {Ivanov}}]{silva_topological_2019}%
  \BibitemOpen
  \bibfield  {author} {\bibinfo {author} {\bibfnamefont {R.~E.~F.}\
  \bibnamefont {Silva}}, \bibinfo {author} {\bibfnamefont {A.}~\bibnamefont
  {Jim\'{e}nez-Gal\'{a}n}}, \bibinfo {author} {\bibfnamefont {B.}~\bibnamefont
  {Amorim}}, \bibinfo {author} {\bibfnamefont {O.}~\bibnamefont {Smirnova}},\
  and\ \bibinfo {author} {\bibfnamefont {M.}~\bibnamefont {Ivanov}},\
  }\bibfield  {title} {\bibinfo {title} {Topological strong-field physics on
  sub-laser-cycle timescale},\ }\href
  {https://doi.org/10.1038/s41566-019-0516-1} {\bibfield  {journal} {\bibinfo
  {journal} {Nat. Photonics}\ }\textbf {\bibinfo {volume} {13}},\ \bibinfo
  {pages} {849} (\bibinfo {year} {2019})}\BibitemShut {NoStop}%
\bibitem [{\citenamefont {Asendorf}(1957)}]{Asendorf1957}%
  \BibitemOpen
  \bibfield  {author} {\bibinfo {author} {\bibfnamefont {R.~H.}\ \bibnamefont
  {Asendorf}},\ }\bibfield  {title} {\bibinfo {title} {Space group of tellurium
  and selenium},\ }\href@noop {} {\bibfield  {journal} {\bibinfo  {journal} {J.
  Chem. Phys.}\ }\textbf {\bibinfo {volume} {27}},\ \bibinfo {pages} {11}
  (\bibinfo {year} {1957})}\BibitemShut {NoStop}%
\bibitem [{\citenamefont {Peng}\ \emph {et~al.}(2015)\citenamefont {Peng},
  \citenamefont {Kioussis},\ and\ \citenamefont {Stewart}}]{Peng2015}%
  \BibitemOpen
  \bibfield  {author} {\bibinfo {author} {\bibfnamefont {H.}~\bibnamefont
  {Peng}}, \bibinfo {author} {\bibfnamefont {N.}~\bibnamefont {Kioussis}},\
  and\ \bibinfo {author} {\bibfnamefont {D.~A.}\ \bibnamefont {Stewart}},\
  }\bibfield  {title} {\bibinfo {title} {Anisotropic lattice thermal
  conductivity in chiral tellurium from first principles},\ }\href@noop {}
  {\bibfield  {journal} {\bibinfo  {journal} {Appl. Phys. Lett.}\ }\textbf
  {\bibinfo {volume} {107}},\ \bibinfo {pages} {251904} (\bibinfo {year}
  {2015})}\BibitemShut {NoStop}%
\bibitem [{\citenamefont {Clark}\ \emph {et~al.}(2005)\citenamefont {Clark},
  \citenamefont {Segall}, \citenamefont {Pickard}, \citenamefont {Hasnip},
  \citenamefont {Probert}, \citenamefont {Refson},\ and\ \citenamefont
  {Payne}}]{castep2005}%
  \BibitemOpen
  \bibfield  {author} {\bibinfo {author} {\bibfnamefont {S.~J.}\ \bibnamefont
  {Clark}}, \bibinfo {author} {\bibfnamefont {M.~D.}\ \bibnamefont {Segall}},
  \bibinfo {author} {\bibfnamefont {C.~J.}\ \bibnamefont {Pickard}}, \bibinfo
  {author} {\bibfnamefont {P.~J.}\ \bibnamefont {Hasnip}}, \bibinfo {author}
  {\bibfnamefont {M.~J.}\ \bibnamefont {Probert}}, \bibinfo {author}
  {\bibfnamefont {K.}~\bibnamefont {Refson}},\ and\ \bibinfo {author}
  {\bibfnamefont {M.~C.}\ \bibnamefont {Payne}},\ }\bibfield  {title} {\bibinfo
  {title} {First principles methods using castep},\ }\href@noop {} {\bibfield
  {journal} {\bibinfo  {journal} {Zeitschrift f\"ur Kristallographie}\ }\textbf
  {\bibinfo {volume} {220}},\ \bibinfo {pages} {567} (\bibinfo {year}
  {2005})}\BibitemShut {NoStop}%
\bibitem [{\citenamefont {Perdew}\ \emph {et~al.}(1996)\citenamefont {Perdew},
  \citenamefont {Burke},\ and\ \citenamefont {Ernzerhof}}]{Perdew1996}%
  \BibitemOpen
  \bibfield  {author} {\bibinfo {author} {\bibfnamefont {J.~P.}\ \bibnamefont
  {Perdew}}, \bibinfo {author} {\bibfnamefont {K.}~\bibnamefont {Burke}},\ and\
  \bibinfo {author} {\bibfnamefont {M.}~\bibnamefont {Ernzerhof}},\ }\bibfield
  {title} {\bibinfo {title} {Generalized gradient approximation made simple},\
  }\href@noop {} {\bibfield  {journal} {\bibinfo  {journal} {Phys. Rev. Lett.}\
  }\textbf {\bibinfo {volume} {77}},\ \bibinfo {pages} {3865} (\bibinfo {year}
  {1996})}\BibitemShut {NoStop}%
\bibitem [{\citenamefont {Tkatchenko}\ and\ \citenamefont
  {Scheffler}(2009)}]{Tkatchenko2009}%
  \BibitemOpen
  \bibfield  {author} {\bibinfo {author} {\bibfnamefont {A.}~\bibnamefont
  {Tkatchenko}}\ and\ \bibinfo {author} {\bibfnamefont {M.}~\bibnamefont
  {Scheffler}},\ }\bibfield  {title} {\bibinfo {title} {Accurate molecular van
  der waals interactions from ground-state electron density and free-atom
  reference data},\ }\href {https://doi.org/10.1103/PhysRevLett.102.073005}
  {\bibfield  {journal} {\bibinfo  {journal} {Phys. Rev. Lett.}\ }\textbf
  {\bibinfo {volume} {102}},\ \bibinfo {pages} {073005} (\bibinfo {year}
  {2009})}\BibitemShut {NoStop}%
\bibitem [{\citenamefont {Andrade}\ \emph {et~al.}(2015)\citenamefont
  {Andrade}, \citenamefont {Strubbe}, \citenamefont {Giovannini}, \citenamefont
  {Larsen}, \citenamefont {Oliveira}, \citenamefont {Alberdi-Rodriguez},
  \citenamefont {Varas}, \citenamefont {Theophilou}, \citenamefont {Helbig},
  \citenamefont {Verstraete}, \citenamefont {Stella}, \citenamefont {Nogueira},
  \citenamefont {Aspuru-Guzik}, \citenamefont {Castro}, \citenamefont
  {Marques},\ and\ \citenamefont {Rubio}}]{Andrade2015}%
  \BibitemOpen
  \bibfield  {author} {\bibinfo {author} {\bibfnamefont {X.}~\bibnamefont
  {Andrade}}, \bibinfo {author} {\bibfnamefont {D.~A.}\ \bibnamefont
  {Strubbe}}, \bibinfo {author} {\bibfnamefont {U.~D.}\ \bibnamefont
  {Giovannini}}, \bibinfo {author} {\bibfnamefont {A.~H.}\ \bibnamefont
  {Larsen}}, \bibinfo {author} {\bibfnamefont {M.~J.~T.}\ \bibnamefont
  {Oliveira}}, \bibinfo {author} {\bibfnamefont {J.}~\bibnamefont
  {Alberdi-Rodriguez}}, \bibinfo {author} {\bibfnamefont {A.}~\bibnamefont
  {Varas}}, \bibinfo {author} {\bibfnamefont {I.}~\bibnamefont {Theophilou}},
  \bibinfo {author} {\bibfnamefont {N.}~\bibnamefont {Helbig}}, \bibinfo
  {author} {\bibfnamefont {M.}~\bibnamefont {Verstraete}}, \bibinfo {author}
  {\bibfnamefont {L.}~\bibnamefont {Stella}}, \bibinfo {author} {\bibfnamefont
  {F.}~\bibnamefont {Nogueira}}, \bibinfo {author} {\bibfnamefont
  {A.}~\bibnamefont {Aspuru-Guzik}}, \bibinfo {author} {\bibfnamefont
  {A.}~\bibnamefont {Castro}}, \bibinfo {author} {\bibfnamefont {M.~A.~L.}\
  \bibnamefont {Marques}},\ and\ \bibinfo {author} {\bibfnamefont
  {A.}~\bibnamefont {Rubio}},\ }\bibfield  {title} {\bibinfo {title}
  {Real-space grids and the octopus code as tools for the development of new
  simulation approaches for electronic systems},\ }\href@noop {} {\bibfield
  {journal} {\bibinfo  {journal} {Phys. Chem. Chem. Phys.}\ }\textbf {\bibinfo
  {volume} {17}},\ \bibinfo {pages} {31371} (\bibinfo {year}
  {2015})}\BibitemShut {NoStop}%
\bibitem [{\citenamefont {Schlipf}\ and\ \citenamefont
  {Gygi}(2015)}]{schlipf_optimization_2015}%
  \BibitemOpen
  \bibfield  {author} {\bibinfo {author} {\bibfnamefont {M.}~\bibnamefont
  {Schlipf}}\ and\ \bibinfo {author} {\bibfnamefont {F.}~\bibnamefont {Gygi}},\
  }\bibfield  {title} {\bibinfo {title} {Optimization algorithm for the
  generation of {ONCV} pseudopotentials},\ }\href
  {https://doi.org/10.1016/j.cpc.2015.05.011} {\bibfield  {journal} {\bibinfo
  {journal} {Comput. Phys. Commun.}\ }\textbf {\bibinfo {volume} {196}},\
  \bibinfo {pages} {36} (\bibinfo {year} {2015})}\BibitemShut {NoStop}%
\bibitem [{\citenamefont {Saito}\ \emph {et~al.}(2017)\citenamefont {Saito},
  \citenamefont {Xia}, \citenamefont {Lu}, \citenamefont {Kanai}, \citenamefont
  {Itatani},\ and\ \citenamefont {Ishii}}]{Saito2017}%
  \BibitemOpen
  \bibfield  {author} {\bibinfo {author} {\bibfnamefont {N.}~\bibnamefont
  {Saito}}, \bibinfo {author} {\bibfnamefont {P.}~\bibnamefont {Xia}}, \bibinfo
  {author} {\bibfnamefont {F.}~\bibnamefont {Lu}}, \bibinfo {author}
  {\bibfnamefont {T.}~\bibnamefont {Kanai}}, \bibinfo {author} {\bibfnamefont
  {J.}~\bibnamefont {Itatani}},\ and\ \bibinfo {author} {\bibfnamefont
  {N.}~\bibnamefont {Ishii}},\ }\bibfield  {title} {\bibinfo {title}
  {Observation of selection rules for circularly polarized fields in
  high-harmonic generation from a crystalline solid},\ }\href@noop {}
  {\bibfield  {journal} {\bibinfo  {journal} {Optica}\ }\textbf {\bibinfo
  {volume} {4}},\ \bibinfo {pages} {1333} (\bibinfo {year} {2017})}\BibitemShut
  {NoStop}%
\bibitem [{\citenamefont {Chen}(2018)}]{Chen2018}%
  \BibitemOpen
  \bibfield  {author} {\bibinfo {author} {\bibfnamefont {Z.-Y.}\ \bibnamefont
  {Chen}},\ }\bibfield  {title} {\bibinfo {title} {Spectral control of high
  harmonics from relativistic plasmas using bicircular fields},\ }\href@noop {}
  {\bibfield  {journal} {\bibinfo  {journal} {Phys. Rev. E}\ }\textbf {\bibinfo
  {volume} {97}},\ \bibinfo {pages} {043202} (\bibinfo {year}
  {2018})}\BibitemShut {NoStop}%
\bibitem [{\citenamefont {Mairesse}\ \emph {et~al.}(2008)\citenamefont
  {Mairesse}, \citenamefont {Dudovich}, \citenamefont {Levesque}, \citenamefont
  {Ivanov}, \citenamefont {Corkum},\ and\ \citenamefont
  {Villeneuve}}]{Mairesse2008}%
  \BibitemOpen
  \bibfield  {author} {\bibinfo {author} {\bibfnamefont {Y.}~\bibnamefont
  {Mairesse}}, \bibinfo {author} {\bibfnamefont {N.}~\bibnamefont {Dudovich}},
  \bibinfo {author} {\bibfnamefont {J.}~\bibnamefont {Levesque}}, \bibinfo
  {author} {\bibfnamefont {M.~Y.}\ \bibnamefont {Ivanov}}, \bibinfo {author}
  {\bibfnamefont {P.~B.}\ \bibnamefont {Corkum}},\ and\ \bibinfo {author}
  {\bibfnamefont {D.~M.}\ \bibnamefont {Villeneuve}},\ }\bibfield  {title}
  {\bibinfo {title} {Electron wavepacket control with elliptically polarized
  laser light in high harmonic generation from aligned molecules},\ }\href@noop
  {} {\bibfield  {journal} {\bibinfo  {journal} {New J. Phys.}\ }\textbf
  {\bibinfo {volume} {10}},\ \bibinfo {pages} {025015} (\bibinfo {year}
  {2008})}\BibitemShut {NoStop}%
\bibitem [{\citenamefont {Wang}\ \emph {et~al.}(2017)\citenamefont {Wang},
  \citenamefont {Zhu}, \citenamefont {Liu}, \citenamefont {Li}, \citenamefont
  {Zhang}, \citenamefont {Lan},\ and\ \citenamefont {Lu}}]{Wang2017}%
  \BibitemOpen
  \bibfield  {author} {\bibinfo {author} {\bibfnamefont {D.}~\bibnamefont
  {Wang}}, \bibinfo {author} {\bibfnamefont {X.}~\bibnamefont {Zhu}}, \bibinfo
  {author} {\bibfnamefont {X.}~\bibnamefont {Liu}}, \bibinfo {author}
  {\bibfnamefont {L.}~\bibnamefont {Li}}, \bibinfo {author} {\bibfnamefont
  {X.}~\bibnamefont {Zhang}}, \bibinfo {author} {\bibfnamefont
  {P.}~\bibnamefont {Lan}},\ and\ \bibinfo {author} {\bibfnamefont
  {P.}~\bibnamefont {Lu}},\ }\bibfield  {title} {\bibinfo {title} {High
  harmonic generation from axial chiral molecules},\ }\href@noop {} {\bibfield
  {journal} {\bibinfo  {journal} {Opt. Express}\ }\textbf {\bibinfo {volume}
  {25}},\ \bibinfo {pages} {23502} (\bibinfo {year} {2017})}\BibitemShut
  {NoStop}%
\bibitem [{\citenamefont {Wu}\ \emph {et~al.}(2018)\citenamefont {Wu},
  \citenamefont {Qiu}, \citenamefont {Wang}, \citenamefont {Wang},\ and\
  \citenamefont {Ye}}]{Wu2018}%
  \BibitemOpen
  \bibfield  {author} {\bibinfo {author} {\bibfnamefont {W.}~\bibnamefont
  {Wu}}, \bibinfo {author} {\bibfnamefont {G.}~\bibnamefont {Qiu}}, \bibinfo
  {author} {\bibfnamefont {Y.}~\bibnamefont {Wang}}, \bibinfo {author}
  {\bibfnamefont {R.}~\bibnamefont {Wang}},\ and\ \bibinfo {author}
  {\bibfnamefont {P.}~\bibnamefont {Ye}},\ }\bibfield  {title} {\bibinfo
  {title} {Tellurene: its physical properties, scalable nanomanufacturing, and
  device applications},\ }\href@noop {} {\bibfield  {journal} {\bibinfo
  {journal} {Chem. Soc. Rev.}\ }\textbf {\bibinfo {volume} {47}},\ \bibinfo
  {pages} {7203} (\bibinfo {year} {2018})}\BibitemShut {NoStop}%
\bibitem [{\citenamefont {Tanaka}\ \emph {et~al.}(2010)\citenamefont {Tanaka},
  \citenamefont {Collins}, \citenamefont {Lovesey}, \citenamefont {Matsumami},
  \citenamefont {Moriwaki},\ and\ \citenamefont {Shin}}]{Tanaka2010}%
  \BibitemOpen
  \bibfield  {author} {\bibinfo {author} {\bibfnamefont {Y.}~\bibnamefont
  {Tanaka}}, \bibinfo {author} {\bibfnamefont {S.~P.}\ \bibnamefont {Collins}},
  \bibinfo {author} {\bibfnamefont {S.~W.}\ \bibnamefont {Lovesey}}, \bibinfo
  {author} {\bibfnamefont {M.}~\bibnamefont {Matsumami}}, \bibinfo {author}
  {\bibfnamefont {T.}~\bibnamefont {Moriwaki}},\ and\ \bibinfo {author}
  {\bibfnamefont {S.}~\bibnamefont {Shin}},\ }\bibfield  {title} {\bibinfo
  {title} {Determination of the absolute chirality of tellurium using resonant
  diffraction with circularly polarized x-rays},\ }\href@noop {} {\bibfield
  {journal} {\bibinfo  {journal} {J. Phys.: Condens. Matter}\ }\textbf
  {\bibinfo {volume} {22}},\ \bibinfo {pages} {122201} (\bibinfo {year}
  {2010})}\BibitemShut {NoStop}%
\bibitem [{\citenamefont {Sakano}\ \emph {et~al.}(2020)\citenamefont {Sakano},
  \citenamefont {Hirayama}, \citenamefont {Takahashi}, \citenamefont {Akebi},
  \citenamefont {Nakayama}, \citenamefont {Kuroda}, \citenamefont {Taguchi},
  \citenamefont {Yoshikawa}, \citenamefont {Miyamoto}, \citenamefont {Okuda},
  \citenamefont {Ono}, \citenamefont {Kumigashira}, \citenamefont {Ideue},
  \citenamefont {Iwasa}, \citenamefont {Mitsuishi}, \citenamefont {Ishizaka},
  \citenamefont {Shin}, \citenamefont {Miyake}, \citenamefont {Murakami},
  \citenamefont {Sasagawa},\ and\ \citenamefont {Kondo}}]{sakano_radial_2020}%
  \BibitemOpen
  \bibfield  {author} {\bibinfo {author} {\bibfnamefont {M.}~\bibnamefont
  {Sakano}}, \bibinfo {author} {\bibfnamefont {M.}~\bibnamefont {Hirayama}},
  \bibinfo {author} {\bibfnamefont {T.}~\bibnamefont {Takahashi}}, \bibinfo
  {author} {\bibfnamefont {S.}~\bibnamefont {Akebi}}, \bibinfo {author}
  {\bibfnamefont {M.}~\bibnamefont {Nakayama}}, \bibinfo {author}
  {\bibfnamefont {K.}~\bibnamefont {Kuroda}}, \bibinfo {author} {\bibfnamefont
  {K.}~\bibnamefont {Taguchi}}, \bibinfo {author} {\bibfnamefont
  {T.}~\bibnamefont {Yoshikawa}}, \bibinfo {author} {\bibfnamefont
  {K.}~\bibnamefont {Miyamoto}}, \bibinfo {author} {\bibfnamefont
  {T.}~\bibnamefont {Okuda}}, \bibinfo {author} {\bibfnamefont
  {K.}~\bibnamefont {Ono}}, \bibinfo {author} {\bibfnamefont {H.}~\bibnamefont
  {Kumigashira}}, \bibinfo {author} {\bibfnamefont {T.}~\bibnamefont {Ideue}},
  \bibinfo {author} {\bibfnamefont {Y.}~\bibnamefont {Iwasa}}, \bibinfo
  {author} {\bibfnamefont {N.}~\bibnamefont {Mitsuishi}}, \bibinfo {author}
  {\bibfnamefont {K.}~\bibnamefont {Ishizaka}}, \bibinfo {author}
  {\bibfnamefont {S.}~\bibnamefont {Shin}}, \bibinfo {author} {\bibfnamefont
  {T.}~\bibnamefont {Miyake}}, \bibinfo {author} {\bibfnamefont
  {S.}~\bibnamefont {Murakami}}, \bibinfo {author} {\bibfnamefont
  {T.}~\bibnamefont {Sasagawa}},\ and\ \bibinfo {author} {\bibfnamefont
  {T.}~\bibnamefont {Kondo}},\ }\bibfield  {title} {\bibinfo {title} {Radial
  {Spin} {Texture} in {Elemental} {Tellurium} with {Chiral} {Crystal}
  {Structure}},\ }\href {https://doi.org/10.1103/PhysRevLett.124.136404}
  {\bibfield  {journal} {\bibinfo  {journal} {Phys. Rev. Lett.}\ }\textbf
  {\bibinfo {volume} {124}},\ \bibinfo {pages} {136404} (\bibinfo {year}
  {2020})}\BibitemShut {NoStop}%
\bibitem [{\citenamefont {Rodriguez}\ \emph {et~al.}(2020)\citenamefont
  {Rodriguez}, \citenamefont {Tsirlin}, \citenamefont {Biesner}, \citenamefont
  {Ueno}, \citenamefont {Takahashi}, \citenamefont {Kobayashi}, \citenamefont
  {Dressel},\ and\ \citenamefont {Uykur}}]{rodriguez_two_2020}%
  \BibitemOpen
  \bibfield  {author} {\bibinfo {author} {\bibfnamefont {D.}~\bibnamefont
  {Rodriguez}}, \bibinfo {author} {\bibfnamefont {A.~A.}\ \bibnamefont
  {Tsirlin}}, \bibinfo {author} {\bibfnamefont {T.}~\bibnamefont {Biesner}},
  \bibinfo {author} {\bibfnamefont {T.}~\bibnamefont {Ueno}}, \bibinfo {author}
  {\bibfnamefont {T.}~\bibnamefont {Takahashi}}, \bibinfo {author}
  {\bibfnamefont {K.}~\bibnamefont {Kobayashi}}, \bibinfo {author}
  {\bibfnamefont {M.}~\bibnamefont {Dressel}},\ and\ \bibinfo {author}
  {\bibfnamefont {E.}~\bibnamefont {Uykur}},\ }\bibfield  {title} {\bibinfo
  {title} {Two {Linear} {Regimes} in {Optical} {Conductivity} of a {Type}-{I}
  {Weyl} {Semimetal}: {The} {Case} of {Elemental} {Tellurium}},\ }\href
  {https://doi.org/10.1103/PhysRevLett.124.136402} {\bibfield  {journal}
  {\bibinfo  {journal} {Phys. Rev. Lett.}\ }\textbf {\bibinfo {volume} {124}},\
  \bibinfo {pages} {136402} (\bibinfo {year} {2020})}\BibitemShut {NoStop}%
\bibitem [{\citenamefont {Xiang}\ \emph {et~al.}(2019)\citenamefont {Xiang},
  \citenamefont {Gao}, \citenamefont {Xua}, \citenamefont {Wu},\ and\
  \citenamefont {Leng}}]{Xiang2019}%
  \BibitemOpen
  \bibfield  {author} {\bibinfo {author} {\bibfnamefont {Y.}~\bibnamefont
  {Xiang}}, \bibinfo {author} {\bibfnamefont {S.}~\bibnamefont {Gao}}, \bibinfo
  {author} {\bibfnamefont {R.-G.}\ \bibnamefont {Xua}}, \bibinfo {author}
  {\bibfnamefont {W.}~\bibnamefont {Wu}},\ and\ \bibinfo {author}
  {\bibfnamefont {Y.}~\bibnamefont {Leng}},\ }\bibfield  {title} {\bibinfo
  {title} {Phase transition in two-dimensional tellurene under mechanical
  strain modulation},\ }\href@noop {} {\bibfield  {journal} {\bibinfo
  {journal} {Nano Energy}\ }\textbf {\bibinfo {volume} {58}},\ \bibinfo {pages}
  {202} (\bibinfo {year} {2019})}\BibitemShut {NoStop}%
\end{thebibliography}%
\end{document}